\documentclass[12pt]{article}
\usepackage{setspace}

\usepackage{graphicx}
\graphicspath{{Figures/}}
\usepackage{amsfonts}
\usepackage{amssymb}

\usepackage{amsmath}
\usepackage[colorlinks,linkcolor=blue,citecolor=blue,urlcolor=blue]{hyperref}  
\usepackage{lineno}

\usepackage{xspace}
\usepackage{array}
\usepackage{caption}
\usepackage{subfig}
\usepackage{cite}
\usepackage{multirow}
\usepackage{enumerate}
\usepackage{mathrsfs}
\usepackage{pdflscape}
\usepackage[utf8]{inputenc}
\usepackage{rotating}
\usepackage{empheq}
\usepackage{standalone}
\usepackage{geometry}
\usepackage{listings}
\usepackage{xcolor}

\usepackage[ruled,vlined]{algorithm2e}
\usepackage{mathtools, nccmath}
\usepackage{booktabs}
\usepackage{epstopdf} 

\usepackage{bm}   

\usepackage[symbol]{footmisc} 

\usepackage{mathdots}  

\usepackage{titlesec} 

\definecolor{codegreen}{rgb}{0,0.6,0}
\definecolor{codegray}{rgb}{0.5,0.5,0.5}
\definecolor{codepurple}{rgb}{0.58,0,0.82}
\definecolor{backcolour}{rgb}{1.0,1.0,0.7}

\lstdefinestyle{mystyle}{
	backgroundcolor=\color{backcolour},   
	commentstyle=\color{codegreen},
	keywordstyle= \color{magenta}, 
	numberstyle=\tiny\color{black},
	stringstyle=\color{codepurple},
	basicstyle=\footnotesize,
	breakatwhitespace=false,         
	breaklines=true,                 
	captionpos=tl,                    
	keepspaces=true,                 
	numbers=left,                    
	numbersep=5pt,                  
	showspaces=false,                
	showstringspaces=false,
	showtabs=false,                  
	tabsize=3,
	basewidth = {.55em}
}

\begin{document}

	
		
\begin{center}
	{\Large  A co-rotational formulation for arbitrarily shaped planar beams under large displacements and rotations}
\end{center}		
		
\centerline{  
    {\bf Linh T. M. Phi} \textsuperscript{a}, 
	{\bf Zhangxian Yuan} \textsuperscript{a,}\footnotemark[1]
}

\noindent\textsuperscript{a}{
	Aerospace Engineering Department, Worcester Polytechnic Institute, Worcester, MA, 01609-2280, USA
}

\footnotetext[1]{\noindent{Corresponding author. Email: zyuan2@wpi.edu } }

\bigskip
{\bf Abstract}			

A planar co-rotational beam formulation is proposed for the large displacement and large rotation analysis of beams with arbitrary initial geometries.
A local frame is attached to each differential beam segment and follows its rigid body translation and rotation, enabling a consistent kinematic description suitable for higher-order basis functions.
An auxiliary straight reference configuration is introduced so that both the arbitrary initial configuration and the current deformed configuration can be described relative to a common reference, enabling arbitrary initial geometries to be treated within the co-rotational framework.
While the framework is general and applicable to a wide range of higher-order basis functions, Non-Uniform Rational B-Splines (NURBS) basis functions are adopted in this study to exploit their superior geometric representation and smoothness.
The performance of the proposed formulation is evaluated through five numerical examples, including static, post-buckling, dynamic, and coupled static-dynamic cases.
The results demonstrate excellent convergence, accurate prediction of large-deformation responses, effective mitigation of locking without additional treatment, and robust performance for beams with non-uniform cross-sections and spatially varying curvature.


	

\section{Introduction}

\indent \indent
Curved beam structures are widely employed in modern engineering design due to their high load-transfer efficiency, geometric flexibility, and aesthetic appeal.
They are particularly attractive for lightweight and high-performance systems, and are commonly found in applications such as arches in civil infrastructure, curved frames in aerospace and automotive engineering, flexible robotic arms, and biomedical devices such as stents and guide-wires.
In many of these applications, curved beams are subjected to large deformations and rotations, accompanied by significant geometric nonlinearities.
Understanding such behavior is of great importance, as it is essential for the design and optimization of structures with desired mechanical performance.

Accurately modeling and capturing this behavior presents significant numerical challenges, including the proper treatment of finite rotations and the avoidance of spurious stiffening effects, such as membrane and shear locking.
When the initial geometry is curved or arbitrarily shaped, additional complexities arise in ensuring accurate geometric representation \cite{Friedman1998,borkovic_rotation-free_2018}.
Therefore, the development of advanced curved beam formulations capable of handling arbitrary initial geometries with high computational accuracy and efficiency remains an active area of research \cite{Weeger2022, Wang2025, jangravi_large_2026}.
This is particularly driven by the increasing demand for real-time simulations in robotics and control \cite{Danesh2025}.

In the literature, a significant amount of effort has been devoted to the modeling of beam structures undergoing large displacements and rotations. 
Representative formulations include the absolute nodal coordinate formulation \cite{Shabana1998, Li2023}, the geometrically exact beam formulations \cite{Zhang2013, Weeger2022}, and the co-rotational formulation \cite{Crisfield1996, Albino2018, battini_co-rotational_2002 }.
The absolute nodal coordinate formulation, originally proposed by Shabana \cite{Shabana1996}, describes the kinematics of a flexible body using positions and slope vectors.
The geometrically exact beam formulation, pioneered by Reissner \cite{Reissner1972} and later extended by Simo and his colleague \cite{Simo1986, Simo1988}, explicitly incorporates both positions and rotations as primary variables.
A comprehensive comparison of absolute nodal coordinate formulation and geometrically exact beam formulations, highlighting their similarities and differences, was reviewed by Romero \cite{Romero2028}.
Co-rotational formulation, initiated by the early work by Wempner \cite{Wempner1969}, Belytschko and his colleagues \cite{Belytschko1977, Belytschko1979}, decomposes the motion into a rigid body movement and a pure local deformation.
In this framework, a local frame is attached to the element and follows the rigid-body translation and rotation, thereby simplifying the treatment of large rotations while maintaining computational efficiency.
Therefore, co-rotational methods have attracted considerable attention, leading to the development of various elements for beams \cite{hsiao_consistent_1999,norachan_co-rotational_2012}, shells  \cite{battini_choice_2006,jiang_co-rotational_1994, felippa_unified_2005}, and solids \cite{Crisfield1996, Cho2018}.

When applied to slender beams, these formulations may suffer from numerical issues such as shear locking \cite{urthaler_corotational_2005,nachbagauer_new_2011,zheng_series_2025, jangravi_large_2026}. 
Consequently, considerable effort has been devoted to the development of locking-free elements \cite{urthaler_corotational_2005,nachbagauer_new_2011,zheng_series_2025, jangravi_large_2026, Magisano2021, Armero2024}, often requiring specialized treatments, such as mixed formulation \cite{garcea_mixed_1998,bai_alleviation_2023 }, reduced integration \cite{noor_mixed_1981,garcia-vallejo_new_2007,Romero2028}, etc.
Alternatively, since the Reissner-Simo geometrically exact beam model accounts for transverse shear deformation, efforts have been made to develop shear-rigid (Bernoulli-Euler-type) geometrically exact beam models for slender beams to avoid shear locking \cite{Zhang2013}.
However, this approach leads to the absence of a unified formulation applicable to both thick and slender beams, and special care is required when selecting appropriate models depending on the beam slenderness.

While the aforementioned formulations provide a robust foundation for nonlinear analysis of beam-type structures, the accurate representation of curved geometries remains a fundamental challenge.
In conventional finite element implementations, curved beams are typically approximated using a sequence of straight beam segments.
Although this piecewise-linear approach is simple and widely adopted, it introduces geometric approximation errors that cannot be completely eliminated through mesh refinement.
While specialized curved beam elements have been developed to alleviate this issue, most existing formulations rely on lower-order basis functions \cite{ crisfield_consistent_1990, galvanetto_energy-conserving_1996, hsiao_consistent_1999, mo_hsiao_co-rotational_2000,battini_co-rotational_2002 }, which typically result in a constant curvature within each element.
As a result, for beams with non-uniform or complex curvature, a large number of elements is still required to achieve sufficient geometric accuracy.
It is worth noting that while curved beam formulations have been extensively studied within geometrically exact formulation and the absolute nodal coordinate formulation, only a limited number of investigations have been carried out using the corotational formulation.
Sandhu et al. \cite{Sandhu1990} and Wang et al. \cite{Wang2025} proposed spatial corotational curved beam elements.
Deng et al. \cite{Deng2023} proposed a planar co-rotational viscoelastic curved beam element.

The majority of existing nonlinear beam formulations are traditionally implemented using low-order finite elements, such as linear elements\cite{battini_co-rotational_2002, Albino2018, crisfield_consistent_1990, mo_hsiao_co-rotational_2000}.
While such discretizations are simple and robust, they often suffer from general limitations of lower-order elements, such as limited accuracy, slow convergence, and numerical issues such as locking, especially for slender structures.
The limitations have motivated increasing interest in higher-order element formulations, which generally offer superior convergence rates and enable more accurate representation of complex deformation states with fewer degrees of freedom (DOF).
Higher-order formulations have been widely used in modeling complex structures, such as functionally graded materials\cite{Wang2017_FGM,Wang2019}, 
stiffened plates\cite{Deng2019}, elliptic interface problem \cite{Zheng2026}, multiscale brain injury modeling \cite{Molchanov2025}.
However, extending existing nonlinear beam formulations from lower-order to higher-order elements is not straightforward.
In co-rotational approaches, an element-wise local frame is typically attached to each element \cite{Crisfield1996,battini_choice_2006, Cho2018, Rankin1986, jiang_co-rotational_1994, felippa_unified_2005} to separate rigid-body motion from deformation.
As the basis function order increases, the kinematic description spans the entire element domain, and the assumption of a single rigid-body rotation for the entire element becomes insufficient to represent the actual motion.
A consistent treatment of rotations over the element domain is therefore required.
To address this limitation, Yuan et al. \cite{yuan_co-rotational_2019} proposed a high-order co-rotational formulation, demonstrating improved accuracy and exponential convergence behavior.
However, that framework was formulated for beams with straight configurations, which limits its ability to model structures with complex initial geometries.
Although one could approximate complex shapes using multiple straight high-order elements, such an approach contradicts the fundamental motivation for using high-order methods, and reintroduces geometric approximation errors similar to those encountered in conventional low-order elements.
Therefore, a gap remains in the development of formulations that can simultaneously achieve accurate representation of arbitrarily shaped geometries and high-order approximation within the co-rotational framework.

A variety of high-order element methods have been developed in the literature.
Among them, spectral element method (SEM) \cite{Seriani1994}, the weak-form quadrature element method (QEM) \cite{Wang2017_QEM}, and isogeometric analysis (IGA) \cite{Hughes2005, NGUYEN201589}, are widely used.
These methods utilize diverse basis functions, such as Lagrange polynomials, Legendre polynomials, and Non-Uniform Rational B-Splines (NURBS).
Pairing with appropriately selected nodal distributions and quadrature schemes, high-order element methods offer several important advantages, including improved accuracy and convergence, reduced computational cost for a given level of accuracy, enhanced capability for representing complex geometries, reduced numerical dispersion in wave propagation problems, and alleviation of locking phenomena \cite{Duczek2018}.
Among these, NURBS-based isogeometric analysis has received particular attention due to its ability to provide an exact geometric representation. 
NURBS, as a generalization of B-splines, are widely used in computer-aided design (CAD) for the smooth and accurate description of complex geometries. 
IGA adopts the same NURBS basis functions to represent both the geometry and the solution fields, thereby establishing a direct link between CAD and analysis. 
This feature makes IGA especially attractive for problems involving curved geometries and high-order continuity requirements.

The objective of this work is to develop a high-order co-rotational formulation for arbitrarily shaped beam elements.
The proposed approach aims to accurately represent complex initial geometries while retaining the efficiency and robustness characteristic of the co-rotational framework.
In traditional IGA, initial shapes are described using NURBS curves, and the physical domain is mapped onto a parametric domain to represent complex geometries.
To ensure our formulation remains general and suitable for various basis functions beyond NURBS, we introduce an auxiliary reference configuration, and both the initial and current configurations are measured with respect to this common auxiliary reference.
This strategy provides a unified approach that can be adapted to different high-order implementations.
To demonstrate the efficacy and accuracy of this approach, NURBS basis functions are adopted in this work to represent both the initial geometry and the displacement field, leveraging their superior geometric representation and smoothness properties.

The remainder of this paper is organized as follows. Section~\ref{section:2} presents the co-rotational formulation for higher-order beam elements.
Section~\ref{section:3} briefly introduces the NURBS basis functions and describes their implementation within the proposed framework. Numerical examples are presented in Section~\ref{section:4} to demonstrate the accuracy and efficiency of the proposed formulation. Finally, conclusions are drawn in Section~\ref{section:5}.

\section{Co-rotational formulation for beams with arbitrary initial shape}
\label{section:2}

\indent
\subsection{Beam kinematics}

\begin{figure}
		\centering
		\includegraphics[width=6in]{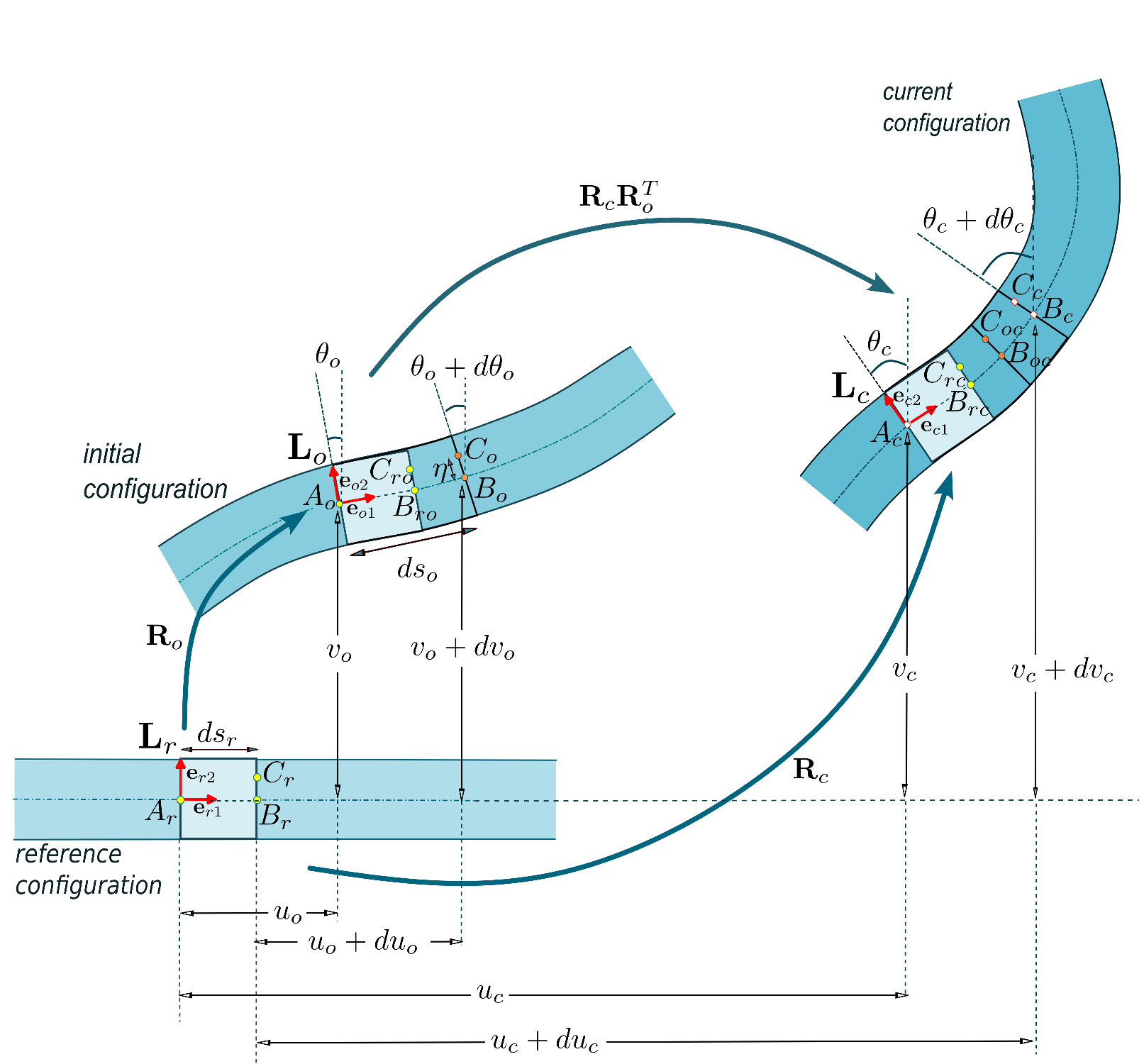}
		\caption{Schematic of the co-rotational formulation for an arbitrarily shaped beam.} 
		\label{fig:frame}
	\end{figure}

    \indent \indent
    For an initially straight co-rotational multi-node high-order beam element, two local frames, one attached to the initial configuration and the other to the current configuration, are used to describe the kinematics of beam deformation \cite{yuan_co-rotational_2019}. 
    When considering an arbitrarily shaped multi-node beam element, in addition to the initial and deformed configurations, an auxiliary straight beam configuration is introduced as a reference configuration. 
    Accordingly, three local Cartesian coordinate systems are defined and shown in Fig.\ \ref{fig:frame}.
    In the initial configuration, a material point $A_o$ is located at the centroid of the beam, with  $s_o$ denoting the arc-length parameter along the centroidal axis.
    Point $A_o$ is mapped to point $A_r$ in the reference configuration and moves to point $A_c$ in the current (deformed) configuration. 
    The arc-length parameters of $A_r$ and $A_c$ in their respective configurations are denoted by $s_r$ and $s_c$.
    Consider a differential beam element of length $ds_o$ in the initial configuration.
    The start and end points of the centroidal axis of this element are $A_o$ and $B_o$, respectively. 
    Point $C_o$ lies on the cross-section passing through $B_o$, at a distance $\eta$ from $B_o$.
	The initial local frame ${\bf L}_o$ is attached to the plane of the cross-section passing through point $A_o$. 
	The basis vector ${\bf e}_{o2}$ lies along the cross-section through $A_o$, and ${\bf e}_{o1}$ is perpendicular to ${\bf e}_{o2}$. 
	After deformation, points $A_o$, $B_o$ and $C_o$ move to points $A_c$, $B_c$ and $C_c$,  respectively.
	The local frame ${\bf L}_o$ co-rotates and translates with the differential element, and is denoted as ${\bf L}_c$ in the current configuration.
	Therefore, ${\bf e}_{o1}$ and ${\bf e}_{c1}$ are the normal directions of the cross-section in the initial and deformed configuration.
	Points $A_r$, $B_r$, and $C_r$ in the reference configuration correspond to $A_o$, $B_o$ and $C_o$.
	The reference frame is attached to the cross-section at point $A_r$.
	Since the reference configuration is a straight beam, ${\bf e}_{r1}$ points from $A_r$ to $B_r$.

	The kinematic relationship is formulated in the reference frame ${\bf L}_r$.
	The rotation angles of the initial and current frames relative to the reference frame ${\bf L}_r$ are denoted by $\theta_o$ and $\theta_c$, respectively.
	The basis vectors of ${\bf L}_o$ and ${\bf L}_c$ frames are given by,
	\begin{equation}
		{\bf e}_{o1} = 
		\begin{bmatrix}
			\cos \theta_o \\
			\sin \theta_o
		\end{bmatrix}  , \quad
		{\bf e}_{o2} = 
		\begin{bmatrix}
			- \sin \theta_o \\
			\cos \theta_o 
		\end{bmatrix}, \quad	
		{\bf e}_{c1} = 
		\begin{bmatrix}
			\cos \theta_c \\
			\sin \theta_c
		\end{bmatrix}  , \quad
		{\bf e}_{c2} = 
		\begin{bmatrix}
			- \sin \theta_c \\
			\cos \theta_c
		\end{bmatrix}.
	\end{equation}
	The transformation matrices from the reference frame ${\bf L}_r$ to the initial ${\bf L}_o$ and current frames ${\bf L}_c$, denoted by ${\bf R}_o$ and ${\bf R}_c$, are,
	\begin{equation}
      {\bf R}_o = 
      \begin{bmatrix}
      	 {\bf e}_{o1} & {\bf e}_{o2}
      \end{bmatrix}  , \quad
       {\bf R}_c = 
        \begin{bmatrix}
       	{\bf e}_{c1} & {\bf e}_{c2} 
       \end{bmatrix}. 
	\end{equation}

	The coordinates of $A_r$ are $(x_r, y_r)$. 
	The corresponding point in the initial configuration, $A_o$, has coordinates $(x_o, y_o)=(x_r + u_o, y_r + v_o)$, and the point in current configuration, $A_c$, has coordinates $(x_c, y_c) = (x_r + u_c, y_r + v_c)$.	
	$u_o$ and $v_o$ denote the components along of ${\bf e}_{r1}$ and ${\bf e}_{r2}$, respectively, of the vector from $A_r$ and $A_o$, while $u_c$ and $v_c$ denote the corresponding components from $A_r$ to $A_c$.
	Therefore, the displacement vector during deformation is,
	\begin{equation}
		u~{\bf e}_{r1} + v~{\bf e}_{r2}  = (u_c-u_o)~{\bf e}_{r1} + (v_c-v_o)~{\bf e}_{r2}.
	\end{equation}
Similarly, the rotation angle during deformation is $\theta=\theta_c - \theta_o$.

The comparison between the initial curved configuration and the auxiliary straight reference configuration is analogous to the comparison between the deformed and straight configurations in Ref. \cite{yuan_co-rotational_2019}. 
Thus, the same kinematic construction can be adopted, namely,

		\begin{equation}
		\begin{array}{*{20}{c}}
				{\overrightarrow {{A_r}{B_r}}  = \left\{ {\begin{array}{*{20}{c}}
							{ds_r}\\
							0
					\end{array}} \right\},}&{\overrightarrow {{B_r}{B_o}}  = \left\{ {\begin{array}{*{20}{c}}
							{{u_o} + d{u_o}}\\
							{{v_o} + d{v_o}}
					\end{array}} \right\},}&{\overrightarrow {{B_o}{C_o}}  = \left\{ {\begin{array}{*{20}{c}}
							{ - \eta \sin ({\theta _o} + d{\theta _o})}\\
							{\eta \cos ({\theta _o} + d{\theta _o})}
					\end{array}} \right\}}, \\

			{\overrightarrow {{A_r}{A_o}}  = \left\{ {\begin{array}{*{20}{c}}
						{{u_o}}\\
						{{v_o}}
				\end{array}} \right\},}&{\overrightarrow {{A_o}{B_{ro}}}  = \left\{ {\begin{array}{*{20}{c}}
						{ds_r \cos {\theta _o}}\\
						{ds_r \sin {\theta _o}}
				\end{array}} \right\},}&{\overrightarrow {{B_{ro}}{C_{ro}}}  = \left\{ {\begin{array}{*{20}{c}}
						{ - \eta \sin {\theta _o}}\\
						{\eta \cos {\theta _o}}
				\end{array}} \right\}}  .
		\end{array}
	\end{equation}
Hence, 
	\begin{equation}
		\begin{array}{{rl}}
			{\overrightarrow {{C_{ro}}{C_o}} } & { = \overrightarrow {{A_r}{B_r}}  + \overrightarrow {{B_r}{B_o}}  + \overrightarrow {{B_o}{C_o}}  - \left( {\overrightarrow {{A_r}{A_o}}  + \overrightarrow {{A_o}{B_{ro}}}  + \overrightarrow {{B_{ro}}{C_{ro}}} } \right)}\\
			{} & { = \left\{ {\begin{array}{*{20}{c}}
						{d{u_o} + ds_r - ds_r \cos {\theta _o} - \eta d{\theta _o}\cos {\theta _o}}\\
						{d{v_o} - ds_r \sin {\theta _o} - \eta d{\theta _o}\sin {\theta _o}}
				\end{array}} \right\}} .
		\end{array}
	\end{equation}
Analogy, ${\overrightarrow {{C_{rc}}{C_c}} }$ can be obtained by replacing the subscript $o$ with $c$,
		\begin{equation}
		\label{Eq.09}
			{\overrightarrow {{C_{rc}}C_{c}} } = 
			{ \left\{ 
				{\begin{array}{*{20}{c}}
						{du_c + ds_r - ds_r \cos \theta_d  - \eta d\theta_c \cos \theta_c }\\
						{dv_c - ds_r \sin \theta_c  - \eta d\theta_c \sin \theta_c }
				\end{array}} 
			\right\}}.
	\end{equation}
	
During deformation, the vector $\overrightarrow{C_{ro}C_o}$ co-rotates with the frame ${\bf L}_o$ and is transformed into $\overrightarrow{C_{rc}C_{oc}}$, i.e.,
\begin{equation}
\overrightarrow {{C_{rc}}{C_{oc}}}  = {\bf R}_c {\bf R}_o^T \overrightarrow {{C_{ro}}{C_o}}
\end{equation}

$\overrightarrow {{C_{oc}}C_c}$ represents the displacement of material point $C$ with the rigid-body motion removed, and is given by
	\begin{equation}
		\label{Eq.12}
		\overrightarrow {{C_{oc}}C_c}  = \overrightarrow {{C_{rc}}C_c}  - \overrightarrow {{C_{rc}}{C_{oc}}}.
	\end{equation}
The strain is measured in the local frame ${ {\bf L}_c}$.
Accordingly, the deformation expressed in ${ {\bf L}_c}$ is 
	\begin{equation}
		\begin{Bmatrix}
		{d\bar u}\\
		{d\bar v}
		\end{Bmatrix} _{ {\bf L}_c} = 
	{ {\bf R} }_c^T \overrightarrow {{C_{oc}}{C_{c}}} 
	= { {\bf R}}_c ^T \overrightarrow {{C_{rc}}C_c}  - {\bf R}_o^T \overrightarrow {{C_{ro}}{C_o}}.
\end{equation}
where the subscript ${ {\bf L}_c}$ indicates that the vector components are expressed in the frame ${\bf L}_c$.

The transverse normal strain $\varepsilon_{22}$ vanishes, and the axial normal strain $\varepsilon_{11}$ and shear strain $\gamma_{12}$ are nonzero and are given by:
\begin{subequations}\label{equ:strain}
\begin{equation}
	\varepsilon _{11} = \frac{d\bar u}{ds_o} = 
	\frac{d\bar u}{ds_r}  \frac{ ds_r }{ ds_o }
	= {\bf{\varepsilon }}\left( s_r \right) - \eta \kappa \left( s_r \right),
\end{equation}
\begin{equation}
	\gamma _{12} = {\frac{{d\bar u}}{{d\eta }} + \frac{{d\bar v}}{{d{s_o}}}} =	
	{\frac{{d\bar u}}{{d\eta }} + \frac{{d\bar v}}{{d{s_r}}}\frac{{d{s_r}}}{{d{s_o}}}} 
	= {\bf{\gamma }}\left( s_r \right),
\end{equation}
with 
\begin{equation}
	{\bf{\varepsilon }}\left( s_r \right) = \frac{ ds_r }{ ds_o } 
	\left( {\varepsilon _c} - {\varepsilon_o}
	\right),
\end{equation}
\begin{equation}
	{\bf{\kappa }}\left( s_r \right) = \frac{ ds_r }{ ds_o } 
	 \left( \kappa _c - \kappa _o \right),
\end{equation}
\begin{equation}
	{\bf{\gamma }}\left( s_r \right) = \frac{ ds_r }{ ds_o } 
	\left( \gamma _c - \gamma_o \right),
\end{equation}

	\label{Eq.15}
	\begin{align}
		{\varepsilon _k} =& \frac{{du_k}}{{ds_r}}\cos \theta_k  + \frac{{dv_k}}{{ds_r}}\sin \theta_k  + \cos \theta_k  - 1,\\
		{\gamma_k} =&  - \frac{{d{u}_k}}{{d{s_r}}}\sin {\theta_k} + \frac{{d{v_k}}}{{ds_r}}\cos {\theta_k } - \sin {\theta_k}, \\
		{\kappa_k} =& \frac{{d\theta_k }}{{ds_r}};    
		\qquad \qquad  \qquad \qquad  (k=o \text{ or } c).
	\end{align} 
\end{subequations}
It is seen that the strain measurement is independent of the selection of the reference frame. 

\subsection{Equations of motion}

\indent \indent
The equations of motion are derived from Hamilton's principle:
\begin{equation}\label{equ:Hamilton}
	\int_{t_1}^{t_2} \delta \left( T - U + W \right) dt = 0.
\end{equation}
where $T$ is the kinetic energy, $U$ is the strain energy, and $W$ is the work done by the external loads. 

	The variation of strain energy stored in a beam during elastic deformation is

\begin{equation}
	\label{Eq.31}
		{\delta { U}} { = \int_0^{{L_o}} {\int_{A({s_o})} 
				{\left[ {E({s_o})\varepsilon \left( {{s_o}} \right)\delta \varepsilon  + E({s_o}){\eta ^2}\kappa ({s_o})\delta \kappa  + G({s_o})\gamma ({s_o})\delta \gamma } \right]
			} } d\Omega d{s_o}}.
\end{equation}
where $L_o$ is the length of the beam element in the initial configuration. 
Substituting Eq.\ (\ref{equ:strain}) and transforming the integration from the initial configuration to the reference configuration yields

\begin{equation}\label{equ:strainEnergyVariation}
	{\delta { U}} =
	\int_0^{L_r}
\begin{bmatrix}
	{\delta \varepsilon _{c}}\\
	{\delta \gamma _{c}} \\
	\delta {\kappa _{c}}
\end{bmatrix}^T
\begin{bmatrix}
	EA & 0 & 0 \\
	0 & EI & 0 \\
	0 & 0 & \kappa_s GA
\end{bmatrix}
\begin{bmatrix}
	\varepsilon_c - {\varepsilon_o} \\
	\kappa_c - \kappa_o \\
	\gamma_c - \gamma_o \\
\end{bmatrix}
\frac{d s_r}{d s_o} d s_r.
	\end{equation}
where $L_r$ is the length of  beam of the reference configuration, $A(s_o)=\int_{A(s_o)} d\Omega$ is the cross-section area of the initial configuration,
$I(s_o) = \int_{A(s_o)} \eta ^ 2 d\Omega $  is the second moment of area, $E(s_o)$ is the Young's modulus, and $G(s_o)$ is the shear modulus, and $\kappa_s$ is shear correction factor.

For easy presentation, two auxiliary quantities are introduced,
\begin{equation}
		\begin{array}{lll}
			&R1_c  = u_{c}'\cos {\theta _{c}} + v_{c}'\sin {\theta _{c}} + \cos {\theta _{c}} ,\\
			&R2_c  =  - u_{c}'\sin {\theta _{c}} + v_{c}'\cos {\theta _{c}} - \sin {\theta _{c}},\\
		\end{array}
\end{equation}
where $'$ is the spatial derivative with respect to $s_r$. 
It further has,
\begin{equation}
	\frac{ \partial R1_c } {\partial \theta_c} = R2_c , \qquad  \frac{ \partial R2_c } {\partial \theta_c} = - R1_c,
\end{equation}
	
		\begin{equation} \label{equ:strainVariation}
			\left\{ 
			{\begin{array}{*{20}{c}}
					{\delta \varepsilon _{c}}\\
					{\delta \gamma _{c}} \\
					\delta {\kappa _{c}}
			\end{array}} 
		\right\}    = \left[ 
		{\begin{array}{*{20}{c}}
					{\cos {\theta _{c}}}  &  {\sin {\theta _{c}}}  &  0 & {{R2_{c}}}\\
					{ - \sin {\theta _{c}}} & {\cos {\theta _{c}}} &   0 & { - {R1_{c}}} \\
					 0 & 0 & 1 & 0 
			\end{array}} \right] 		
		{\begin{Bmatrix}
					{\delta u_{c}'}\\
					{\delta v_{c}'}\\
					{\delta \theta _{c}' } \\
					{\delta \theta _{c}}
		\end{Bmatrix}} 
	={\bm \Lambda} 
			{\begin{Bmatrix}
			{\delta u_{c}'}\\
			{\delta v_{c}'}\\
			{\delta \theta _{c}' } \\
			{\delta \theta _{c}}
	\end{Bmatrix}} .
	\end{equation}

		The variation of the work done by external loads, including concentrated and distributed loads, is

\begin{equation}
	\delta {W} = \sum_{i = 1}^{NF} \delta {\bm u}_c (\tilde s_{o_i})^T
	\begin{bmatrix}
		\tilde{P}_i \\
		\tilde Q_i \\
		\tilde M_i
	\end{bmatrix}
	+
	\int_0^{{L_r}} 	\delta {\bm u}_c^T
		\begin{bmatrix}
		\tilde p \\
		\tilde q \\
		\tilde m
	\end{bmatrix}
	\frac{d{s_o}}{d{s_r}}
	d s_r,
	\end{equation}
where $\tilde P_i$, $\tilde Q_i$, and $\tilde M_i$ are the concentrated axial force, shear force, and bending moments applied at $s_o=\tilde s_{o_i}$, ($i=1,...,NF$), respectively. $NF$ is the total number of concentrated loads.
$\tilde p$, $\tilde q$, and $\tilde m$ are the distributed axial force, shear force, and bending moment.

The kinetic energy of the beam is 
\begin{equation}
	T = \int_0^{L_o} \left[ 
	\rho A(s_o) \left( \frac{ \partial ( u_c -u_o )}{\partial t} \right)^2 
	+ \rho A(s_o) \left( \frac{ \partial ( v_c -v_o )}{\partial t} \right)^2 
	+ \rho I_\theta(s_o) \left( \frac{ \partial ( \theta_c -\theta_o )}{\partial t} \right)^2 
	\right] 	ds_o,
\end{equation}
where the overdots stand for the time derivatives with respect to time $t$. The double overdots are the second-order derivatives with respective to time. 
$I_\theta(s_o) = \int_{A(s_o)} \eta ^ 2 d\Omega $.
Thus, the variation of the kinetic energy is,
\begin{equation}
	\delta T = \int_0^{L_r}  \delta {\bm u_c}^T 
	\begin{bmatrix}
		\rho A & 0  & 0\\
		0 & \rho A & 0 \\
		0  & 0 & \rho I_\theta
	\end{bmatrix}
	\ddot{ \bm u}_c  \frac{d{s_o}}{d{s_r}}
	ds_r.
\end{equation}

\section{NURBS-based implementation}
\label{section:3}

\indent \indent
The co-rotational formulation developed in the previous section is independent of the specific choice of basis functions and can be implemented with various higher-order discretizations.
In this study, NURBS are adopted as a representative implementation due to their smoothness, high-order continuity, and superior capability in representing complex geometries.
The resulting discretization follows the main idea of isogeometric analysis, in which the same basis functions are used to represent the geometry and approximate the solution fields.

	\subsection{B-spline and NURBS basis functions}

    \indent \indent
For completeness, B-spline and NURBS basis functions are briefly discussed. 
A B-Spline curve $\bm C(\xi)$ of order $p$ is constructed with $n$ control points,  $\bm P_i $ ($i=1$, 2, ..., $n$), and is defined as:
	\begin{equation}\label{equ:b_spline_curve}
		\bm C(\xi)   =\displaystyle\sum_{i=1}^{n} B_{i,p}(\xi)  \bm P_i,
	\end{equation}
	where $\xi$ is the parametric variable.
	$B_{i,p}(\xi)$ are the B-Spline basis function of order $p$ defined with a knot vector, $\Xi = \{ \xi_1, \xi_2, ...,\xi_i,...,\xi_{n+p+1} \} $, $\xi_i \le \xi_{i+1}$, where $\xi_i$ is the $i$-th knot.

The definition of a knot vector is not unique. 
Among the commonly used choices, the open uniform knot vector is widely adopted in isogeometric analysis and is therefore used in this work. 
In a uniform open knot vector, the first and last knots are repeated $p+1$ times.
For a parametric domain $\xi \in [-1,1]$, it is given by,%
\begin{equation}\label{equ:knot_vector}
	\Xi = \left\{ {\overbrace{-1, -1,..., -1}^{p+1}}, \ 
    { \underbrace{ -1 +\frac {1 \times 2}{n-p},  -1+\frac {2 \times 2}{n-p}, ...,-1+ \frac {(n-p-1) \times 2}{n-p}}_{n-p-1} }, \
    {\overbrace{1, 1,..., 1}^{p+1}} 
    \right\}.
\end{equation}

B-Spline basis functions $B_{i,p}(\xi)$ is calculated by the Cox-de Boor recursion formula, namely
\begin{subequations}\label{equ:basis_functions}
	\begin{equation}
		\begin{aligned}
			B_{i,0}(\xi) =
			\begin{cases}
				1  \qquad \text { if } \quad \xi_i \leq \xi < \xi_{i+1} \\
				0 \qquad \text { otherwise}
			\end{cases} 
		\end{aligned},
	\end{equation}
	\begin{equation}
		B_{i,p}(\xi) =  {\frac {\xi - \xi_i} {\xi_{i+p} - \xi_i}} B_{i,p-1}(\xi) + 
		{\frac {\xi_{i+p+1} - \xi} {\xi_{i+p+1} - \xi_{i+1}}} B_{i+1,p-1}(\xi).
	\end{equation}
\end{subequations}

Since B-splines are polynomial functions, they cannot exactly represent certain geometries, such as circle, conic sections.
To overcome this limitation, Non-Uniform Rational B-Splines (NURBS), which are a rational generalization of B-splines, have been introduced.

A NURBS curve is defined as,
\begin{equation}
	\bm C(\xi) 
	= {\frac {\sum_{i=1}^{n} B_{i,p}(\xi) w_i  }  {\sum_{j=1}^{n} B_{j,p}(\xi) w_j} } {\bm P_i }
	= \displaystyle\sum_{i=1}^{n} N_{i,p}(\xi)  \bm P_i,
\end{equation}
where $w_i$ are the weights associated with the control points.
When all weights are equal to unity, i.e., $w_i=1$, the NURBS basis functions $ N_{i,p}(\xi) = B_{i,p}(\xi) w_i / {\sum_{j=1}^{n} B_{j,p}(\xi) w_j}$ reduce to a B-Spline basis functions.

The $k$-th derivative of the NURBS basis function is given by:
\begin{equation}\label{equ:derivative_basis_functions}
	N_{i,p}^{(k)} = \frac {p!}{(p-k)!} \displaystyle\sum_{j=0}^{k} a_{k,j} N_{i+j,p-k} ,
\end{equation}
with
\begin{equation*}
	\begin{aligned}
		a_{0,0} = & 1, \\ 
		a_{k,0} = &\frac {a_{k-1,0}} {\xi_{i+p-k+1} - \xi_i}, \\
		a_{k,j} = &\frac {a_{k-1,j} - a_{k-1,j-1} } {\xi_{i+p+j-k+1} - \xi_{i+j}}, \qquad \qquad j=1,...,k-1 \\
		a_{k,k} = &\frac {-a_{k-1,k-1}} {\xi_{i+p+1} - \xi_{i+k}}.\\
	\end{aligned}
\end{equation*}
For brevity, the first derivative $N_{i,p}^{(1)}$ is denoted by $N_{i,p}'$ in the following sections.

\subsection{ NURBS representation of the initial and reference configurations}\label{sec:decribeInitialShape}

\indent \indent
To construct the NURBS representation of the initial configuration, sampling points $\bm D_{i}=[D_i^x, D_i^y, D_i^{\theta} ]$ ($i=1,2,...,n_{d}$), with $n_{d}$ being the number of sampling points, are taken from the initial configuration. 
The set of data points is 
\begin{equation}
	[ {\bm D} ] = 
	\begin{bmatrix}
		D^x_{1} & D^y_{1} & D^{\theta}_{1}    \\
		D^x_{2} & D^y_{2} & D^{\theta}_{2}     \\
		\vdots  & \vdots  & \vdots \\
		D^x_{n_{d}} & D^y_{n_{d}} & D^{\theta}_{n_{d}}    
	\end{bmatrix}.
\end{equation}

The initial configuration is typically described in two forms: a parametric representation or non-parametric representations, such as implicit formulations and discrete point sets.
When the initial shape of the beam is given by a parametric function with parameter $\zeta \subset [\zeta_{min}, \zeta_{max}]$ 
the data points can be uniformly sampled over the parametric domain as, 
\begin{equation}
\zeta_i = \zeta_{min} +  (\zeta_{max}-\zeta_{min} ) \frac{i-1}{n_d - 1}, \qquad i = 1, ...,n_d
\end{equation}
The parameter value of the $i$-th data point on the NURBS representation, denoted by $\bar {\xi}_{i}$, is defined as, 
\begin{equation}\label{equ:data_point_parameter_values}
	\bar {\xi}_{i} = 2 \frac{ \zeta_i - \zeta_{min} } {\zeta_{max}-\zeta_{min}} - 1 
	=\frac{2(i-1)}{n_d - 1} - 1
	, \qquad i=1,..., n_{d}
\end{equation}

For the second case, where the initial configuration is defined either implicitly or by a set of discrete data points, the parameter values $\bar {\xi}_{i}$ of the $i$-th data point are determined by the chord length method \cite{imani_nurbs-based_2012, piegl_nurbs_1996} as
\begin{equation}
	\begin{array}{l}
		{\bar \xi _1} =  - 1,\\
		{\bar \xi _i} = {\bar \xi _{i - 1}} + 
		\frac{\left\| {\bm D_i} - {\bm D_{i - 1}} \right\|} 
			 {2\sum_{j = 2}^{{n_d} - 1} {\left\| {\bm D_j - {\bm D_{j - 1}}} \right\|} }.    \qquad i=2,..., n_{d}
	\end{array}
\end{equation}
where the Euclidean norm is defined as $\left\| {\bm D}_i \right\| = \sqrt{  {D_{i}^x}^2 + {D_{i}^y}^2 }$


Given a set of data points $[{\bm D}]$ and the corresponding parametric coordinates ${\bar \xi _i}$, the control points can be determined accordingly. 
To construct a $p$-th order NURBS curve with an open knot vector, the number of control points satisfies $p+1 \le n \le n_d$. 
We enforce the NURBS curve to pass through the first and last data points.
 Accordingly, the first and last control points coincide with the corresponding data points, respectively, namely, $\bm P_{o1} = \bm D_1$, and $\bm P_{on} = \bm D_{n_{d}}$.
The remaining control points, $[ {\bm P_o}] = [\bm P_{o2}, \bm P_{o3},...,\bm P_{o(n-1)} ]^T$,  are then obtained using a least-squares data fitting process, given as,
\begin{subequations}\label{equ:data_fitting}
\begin{equation}
	[ {\bm P_o}] =
	\Big(	[ {\bm N}^{1D}]^{T}  	[ {\bm N}^{1D}] \Big) ^ {-1}
	[ {\bm N}^{1D}]^{T} [ {\bm Q}],
\end{equation}
with
	\begin{equation}	
		[ {\bm N}^{1D}]
		= 
		\begin{bmatrix}
			N_{2,p}(\bar \xi_2) & N_{3,p}(\bar \xi_2) & \cdots & N_{n_{}-1,p}(\bar \xi_2) \\
			N_{2,p}(\bar \xi_3) & N_{3,p}(\bar \xi_3) & \cdots & N_{n_{}-1,p}(\bar \xi_3) \\
			\vdots  & \vdots  & \ddots & \vdots  \\
			N_{2,p}(\bar \xi_{n_{d}-1}) & N_{3,p}(\bar \xi_{n_{d}-1}) & \cdots & N_{n_{}-1,p}(\bar \xi_{n_{d}-1}) \\
		\end{bmatrix},
	\end{equation}
	and 
	\begin{equation}
		[ {\bm Q}] = [ {\bm  Q}_{2},  { \bm Q}_3,..., {\bm  Q}_{n_{d}-1} ]^T,
	\end{equation}
	\begin{equation}
		{\bm  Q}_{i} = {\bm D}_{i} - N_{1,p}(\bar \xi_i) {\bm D}_{1} - N_{n_{},p}(\bar \xi_i) {\bm D}_{n_{d}} 
		\qquad
		(i=2,3,...,n_{d}-1).
	\end{equation}
\end{subequations}
Therefore, the initial configuration is given by,
\begin{equation}\label{equ:NURBSrepresent_initial}
	{{{\bm x}}_o}(\xi )  = \sum_{i = 1}^{{n}} {{N_{i,p}}(\xi ){\bf{P}}_{o_i}} 
	\qquad \text{or} \qquad
	\begin{bmatrix}
		x_o  ( \xi )  \\
		y_o  ( \xi )\\
		\theta_o  ( \xi )
	\end{bmatrix} =  \sum_{i = 1}^{{n}} N_{i,p} ( \xi )
	\begin{bmatrix}
		x_{o_i}\\
		y_{o_i} \\
		\theta_{o_i}
	\end{bmatrix}.
\end{equation}

To construct the reference configuration, each data point selected on the initial configuration is mapped onto the reference configuration. 
The formulation is independent of the specific choice of the straight reference configuration, as shown in Eq.\ (\ref{equ:strain}).
For simplicity, in this study, $n_d$ data points are defined on the reference configuration along the arc-length coordinate $s_r$, following the same distribution as those in the initial configuration, i.e.,
\begin{equation}
		{\bm D}_{r_i}  = [s_{r_i} / 2, 0, 0 ]^T 
		 ,\qquad \text{with} 	\qquad
		 \bar s_{r_i} = L_r ( \bar \xi_i + 1 ) / 2 .
\end{equation}
These data points also share the same parametric coordinates $\bar{\xi}_i$ on the NURBS representation.
Following the same procedure, the locations of the $n$ control points in the reference configuration are obtained. 
The reference configuration is then represented as
\begin{equation}\label{equ:NURBSrepresent_refer}
	{{{\bm x}}_r}(\xi )  = \sum_{i = 1}^{{n}} {{N_{i,p}}(\xi ){\bf{P}}_{r_i}} 
	\qquad \text{or} \qquad
	\begin{bmatrix}
		x_r  ( \xi )  \\
		y_r  ( \xi )\\
		\theta_r  ( \xi )
	\end{bmatrix} =  \sum_{i = 1}^{{n}} N_{i,p} ( \xi )
	\begin{bmatrix}
		x_{r_i}\\
		y_{r_i} \\
		\theta_{r_i}
	\end{bmatrix}.
\end{equation}
Therefore, the displacement field from the reference configuration to the initial configuration is expressed as
\begin{subequations}
\begin{equation}
	{\bm u}_o (\xi )  = \sum_{i = 1}^{n} {N_{i,p}}(\xi )  {\bm u}_{o_i} 
    \qquad \text{or} \qquad
	\begin{bmatrix}
		u_o  ( \xi )  \\
		v_o  ( \xi )\\
		\theta_o  ( \xi )
	\end{bmatrix} =  \sum_{i = 1}^{{n}} N_{i,p} ( \xi )
	\begin{bmatrix}
		u_{o_i}\\
		v_{o_i} \\
		\theta_{o_i}
	\end{bmatrix},
\end{equation}
with 
\begin{equation}
	{\bm u}_{o_i} =  {\bf{P}}_{o_i} - {\bf{P}}_{r_i} .
\end{equation}
\end{subequations}

Furthermore, the arc-length coordinate of the reference configuration $s_r$ is related to the NURBS parametric domain by
\begin{equation}
	s_r=L_r(\xi + 1 )/2.
\end{equation}
The Jacobian between the initial configuration and the reference configuration is given by
\begin{equation}
	\begin{array}{rl}
	\frac{d s_o }{d s_r} (\xi) & =  \sqrt{ 
		\left( \frac{ \partial x_o}{\partial s_r}  \right)^2 +
		\left( \frac{ \partial y_o}{\partial s_r}  \right)^2
	} \\
	& =   \frac{2}{L_r} \sqrt{ 
		 \left( \sum_{i = 1}^{{n}} N_{i,p}^{(1)} ( \xi ) x_{o_i} \right)^2 + 
		 \left( \sum_{i = 1}^{{n}} N_{i,p}^{(1)} ( \xi ) y_{o_i} \right)^2.
	}
	\end{array}
\end{equation}

\subsection{Patch-level formulation}

\indent \indent
For the deformed shape, the same basis functions for initial and reference configurations in Eqs.\ (\ref{equ:NURBSrepresent_initial}) and (\ref{equ:NURBSrepresent_refer}) are also employed to interpolate the generalized displacement field, namely, 
\begin{equation}
	{\bm u}_c (\xi )  = \sum_{i = 1}^{n} {N_{i,p}}(\xi )  {\bm u}_{c_i} \\
	\qquad \text{or} \qquad
	\begin{bmatrix}
		u_c  ( \xi )  \\
		v_c  ( \xi )\\
		\theta_c  ( \xi )
	\end{bmatrix} =  \sum_{i = 1}^{{n}} N_{i,p} ( \xi )
	\begin{bmatrix}
		u_{c_i}\\
		v_{c_i} \\
		\theta_{c_i}
	\end{bmatrix},
\end{equation}
where ${\bm u}_{c_i}$ is the generalized displacement of $i$-th control points. 
The displacement vector of all control points within a single patch is
\begin{equation}
	{\bm u}_{c}^e = 
	\begin{bmatrix}
     {\bm u}_{c_1}^T & {\bm u}_{c_2}^T & \cdots & {\bm u}_{c_i}^T & \cdots & {\bm u}_{c_n}^T
	\end{bmatrix}^T  .
\end{equation}

The variations of $u_c'$, $v_c'$, $\theta_c'$, and $\theta_c$ are expressed in terms of the control point DOFs as,

\begin{equation}
	{\begin{Bmatrix}
			{\delta u_{c}'}\\
			{\delta v_{c}'}\\
			{\delta \theta _{c}' } \\
			{\delta \theta _{c}}
	\end{Bmatrix}}  
= \sum_{i = 1}^{{n}} {\bm B}_i \delta{\bm u}_{c_i} =
	\begin{bmatrix}
	{\bm B}_{1} & {\bm B}_{2} & \cdots & {\bm B}_{i} & \cdots & {\bm B}_{n}
\end{bmatrix} \delta {\bm u}_c^e,
		\end{equation}		
with 
\begin{equation}
	{\bm B}_{i} = 
		\begin{bmatrix}
		\frac{2}{L_r}   N_{i,p}' & 0 &  0 \\
		0 & 	\frac{2}{L_r}   N_{i,p}' &  0 \\
		0 &  0& 	\frac{2}{L_r}   N_{i,p}'  \\
		0 &  0 & 0 &  N_{i,p} \\
	\end{bmatrix}.
\end{equation}
The strain variation is given by
		\begin{equation} \label{equ:strainVariation_CP}
	{\begin{Bmatrix}
			{\delta \varepsilon _{c}}\\
			{\delta \gamma _{c}} \\
			\delta {\kappa _{c}}
	\end{Bmatrix}}    = 
 {\bm \Lambda}
	\begin{bmatrix}
	{\bm B}_{1} & {\bm B}_{2} & \cdots & {\bm B}_{i} & \cdots & {\bm B}_{n}
\end{bmatrix} \delta {\bm u}_c^e.
\end{equation}

Substituting  Eq.\ (\ref{equ:strainVariation_CP}) into Eq.\ (\ref{equ:strainEnergyVariation}) yields the variation of the strain energy,
\begin{equation}
	\delta U = {\delta {\bm u}_c^e}^T {\bm f}^{int} = {\delta {\bm u}_c^e}^T
	\begin{bmatrix}
		{{\bm f}^{int}_1}^T & \cdots & {{\bm f}^{int}_j}^T & \cdots & {{\bm f}^{int}_n}^T \\
	\end{bmatrix}^T,
\end{equation}

with
\begin{equation}
	{\bm f}^{int}_j  =	\int_0^{L_r} {{\bm B}_j} ^T {\bm \Lambda}^T 
    \begin{bmatrix}
	EA & 0 & 0 \\
	0 & EI & 0 \\
	0 & 0 & \kappa_s GA
\end{bmatrix}
	\begin{bmatrix}
		\varepsilon_c - {\varepsilon_o} \\
		\kappa_c - \kappa_o \\
		\gamma_c - \gamma_o \\
	\end{bmatrix}
	\frac{d s_r}{d s_0} d s_r,
\end{equation}
where ${\bm f}^{int}$ is the internal nodal force vector.
The tangent stiffness matrix ${\bm K}_T$ is obtained by linearizing the internal force vector with respect to the control point degrees of freedom,
\begin{equation}\label{equ:tangentStiffMatrix}
	{\bm K}_T  = 
	\frac{\partial {\bm f}^{int}} { \partial {\bm u}^e_c}  = 
	\begin{bmatrix}
		{\bm K}_{T,11} & \cdots & {\bm K}_{T,1k} & \cdots & {\bm K}_{T,1n} \\
		\vdots         & \ddots & \vdots         &        & \vdots         \\
		{\bm K}_{T,j1} & \cdots & {\bm K}_{T,jk} & \cdots & {\bm K}_{T,jn} \\
		\vdots         &        & \vdots         & \ddots & \vdots         \\
		{\bm K}_{T,n1} & \cdots & {\bm K}_{T,nk} & \cdots & {\bm K}_{T,nn}
	\end{bmatrix} .
\end{equation} 
The explicit expression of ${\bm K}_{T, jk}$ is provided in Appendix A.

The variation of the work done by the external forces is,
\begin{equation}
		\delta W = {\delta {\bm u}_c^e}^T {\bm f}^{ext} = {\delta {\bm u}_c^e}^T
	\begin{bmatrix}
		{{\bm f}^{ext}_1}^T & \cdots & {{\bm f}^{ext}_j}^T & \cdots & {{\bm f}^{ext}_n}^T \\
	\end{bmatrix},
\end{equation}
with
\begin{equation}
	{\bm f}^{ext}_j = \sum_{i = 1}^{NF} 
	 N_{j,p} ( \tilde{ \xi } )
	\begin{bmatrix}
		\tilde{P}_i \\
		\tilde{Q}_i \\
		\tilde{M}_i
	\end{bmatrix}
	+
	\int_0^{{L_r}} 	 N_{j,p}
	\begin{bmatrix}
		\tilde p \\
		\tilde q \\
		\tilde m
	\end{bmatrix}
	\frac{d{s_o}}{d{s_r}}
	d s_r.
\end{equation}
where ${\bm f}^{ext}$ is the external force vector.

The variation of the kinetic energy is,
\begin{equation}
	\delta T = {\delta {\bm u}_c^e}^T {\bm M} \ddot{\bm u}_c^e ,
\end{equation}

with
\begin{equation}
	{\bm M} =
	\begin{bmatrix}
		{\bm M}_{11} & \cdots & {\bm M}_{1k} & \cdots &  {\bm M}_{1n} \\
		\vdots       & \ddots & \vdots       & \iddots &  \vdots       \\
		{\bm M}_{j1} & \cdots & {\bm M}_{jk} & \cdots &  {\bm M}_{jn} \\
		\vdots       & \iddots& \vdots       & \ddots &  \vdots       \\
		{\bm M}_{n1} & \cdots & {\bm M}_{nk} & \cdots &  {\bm M}_{nn} \\
	\end{bmatrix} ,
\end{equation}
\begin{equation}\label{equ:EOM_NURBS}
	{\bm M}_{jk} 
	= \int_0^{L_r} N_{j,p}N_{k,p}
	\begin{bmatrix}
		\rho A & 0  & 0\\
		0 & \rho A & 0 \\
		0  & 0 & \rho I_\theta
	\end{bmatrix}
	\frac{{d{s_o}}}{{d{s_r}}}
	ds_r,
\end{equation}
where ${\bm M}$ is the mass matrix

Therefore, using Hamilton's principle, the equation of motion for a patch is obtained as
\begin{equation}\label{equ:EOM}
	{\bm M} \ddot{\bm u}_c^e + {\bm C}_d \dot{\bm u}_c^e + {\bm f^{int}} = {\bm f^{ext}}.
\end{equation}
where ${\bm C}_d$ is the damping matrix. 
If Rayleigh proportional damping is considered, the damping matrix is constructed as the linear combination of the mass matrix and the tangent stiffness matrix.

Each control point possesses three DOFs: displacements along the two spatial axes and one rotational DOF. Thus, for a single patch with $n$ control points, the total number of DOFs is $3n$.
The knot span divides one patch into a set of elements.
Gaussian quadrature is performed over each knot span to evaluate matrices and vectors.
For multi-patch formulations, patches are connected through shared common control points. 
A global coordinate frame is introduced to assemble each patch's matrices and vectors, which are derived in their respective local coordinate systems. 
Assembling these matrices and vectors \cite{Hughes2005,NGUYEN201589} leads to the global matrices of the entire structure, resulting in a formulation analogous to Eq.\ (\ref{equ:EOM_NURBS}).

\subsection{Solution strategies for dynamic and static analysis}

\indent \indent
For the dynamic response, Eq.\ (\ref{equ:EOM}) can be solved with various algorithms and numerical methods.
In this study, the time domain response was obtained by the Newmark method \cite{Subbaraj1989} with average acceleration.
At time $t+\Delta t$, the control point displacement, velocity, and acceleration vectors are denoted by ${}^{t+\Delta t}{\bm u}_c^e$, ${}^{t+\Delta t}\dot{\bm u}_c^e$, and ${}^{t+\Delta t}\ddot{\bm u}_c^e$, respectively. The Newmark update equations are written as
\begin{subequations}
\begin{align}
	{}^{t+\Delta t}{\bm u}_c^e &
	=
	{}^{t}{\bm u}_c^e
	+
	{}^{t}\dot{\bm u}_c^e \Delta t\
	+
	\left[
	\left(\frac{1}{2}-\beta\right){}^{t}\ddot{\bm u}_c^e
	+
	\beta\,{}^{t+\Delta t}\ddot{\bm u}_c^e
	\right] \Delta t^2 , \\
	{}^{t+\Delta t}\dot{\bm u}_c^e &
	=
	{}^{t}\dot{\bm u}_c^e
	+
	\left[
	(1-\gamma){}^{t}\ddot{\bm u}_c^e
	+
	\gamma\,{}^{t+\Delta t}\ddot{\bm u}_c^e
	\right] \Delta t ,
    \end{align}
\end{subequations}
where $\Delta t$ is the time-step size.
In this work, the average acceleration parameters $\beta = 1/4$ and $\gamma = 1/2$ are used.

Considering the equation of motion at time $t+\Delta t$,
\begin{equation}
	{\bm M}\,{}^{t+\Delta t}\ddot{\bm u}_c^e
	+
	{\bm C}_d\,{}^{t+\Delta t}\dot{\bm u}_c^e
	+
	{}^{t+\Delta t}{\bm f}^{int}
	=
	{}^{t+\Delta t}{\bm f}^{ext}.
\end{equation}
Since the internal force vector ${}^{t+\Delta t}{\bm f}^{int}$ depends on the displacement field at the current time step, the nonlinear equation of motion is solved using the Newton--Raphson method. 
At time $t+\Delta t$, the incremental form of the equation of motion is written as
\begin{equation}
\begin{aligned}
&
\left(
{}^{t+\Delta t}{\bm K}_{T}^{(i-1)}
+
\frac{1}{\beta \Delta t^2}{\bm M}
+
\frac{\gamma}{\beta \Delta t}{\bm C}_d
\right)
\Delta {\bm u}_c^{e,(i)}
\\
&=
{}^{t+\Delta t}{\bm f}^{ext}
-
{}^{t+\Delta t}{\bm f}^{int,(i-1)}
-
\left(
\frac{1}{\beta \Delta t^2}{\bm M}
+
\frac{\gamma}{\beta \Delta t}{\bm C}_d
\right)
{}^{t+\Delta t}{\bm u}_c^{e,(i-1)}
\\
&\quad
+
{\bm M}
\left[
\frac{1}{\beta \Delta t^2}\,{}^{t}{\bm u}_c^e
+
\frac{1}{\beta \Delta t}\,{}^{t}\dot{\bm u}_c^e
+
\left(
\frac{1}{2\beta}-1
\right)
{}^{t}\ddot{\bm u}_c^e
\right]
\\
&\quad
+
{\bm C}_d
\left[
\frac{\gamma}{\beta \Delta t}\,{}^{t}{\bm u}_c^e
+
\left(
\frac{\gamma}{\beta}-1
\right)
{}^{t}\dot{\bm u}_c^e
+
\frac{\Delta t}{2}
\left(
\frac{\gamma}{\beta}-2
\right)
{}^{t}\ddot{\bm u}_c^e
\right],
\end{aligned}
\label{equ:newmark_incremental}
\end{equation}
with
\begin{equation}
{}^{t+\Delta t}{\bm u}_c^{e,(i)}
=
{}^{t+\Delta t}{\bm u}_c^{e,(i-1)}
+
\Delta {\bm u}_c^{e,(i)}.
\label{equ:newmark_displacement_update}
\end{equation}
Here, the superscript $(i)$ denotes the Newton--Raphson iteration number.

For static analysis, the inertial and damping terms in Eq.
 (\ref{equ:EOM}) are neglected, leading to the equilibrium equations,
\begin{equation}
	{\bm G} = {\bm f}^{int} - {\bm f}^{ext} = {\bm 0}. 
\end{equation}
In this work, the static equilibrium problem is solved using the Newton--Raphson method. 
For problems involving instability or limit points, the pseudo--arclength continuation method with a branch-switching technique \cite{yuan_nonlinear_2018} is employed to trace the equilibrium path and identify potential limit points.

Once the control point displacement vector, ${\bm u}_c^e$, is obtained, the beam deformation ${\bm u}(\xi) =[u(\xi), v(\xi), \theta(\xi) ]^T$ is given by 
\begin{equation}
	{\bm u}(\xi) = \sum_{i = 1}^{n} {N_{i,p}}(\xi ) {\bm u}_i
	=  \sum_{i = 1}^{n} {N_{i,p}}(\xi ) \left( {\bm u}_{c_i} - {\bm u}_{r_i} \right),
\end{equation}
where ${\bm u}_i=[u_i, v_i, \theta_i]^T$ denotes the displacement vector associated with the $i$-th control point.

\section{Numerical examples}
	\label{section:4}

\indent \indent It is known that some geometries,  such as conics, circles, etc, cannot be represented exactly by a B-spline, such as conics, circles, etc.
Exact representation of such geometries requires NURBS with appropriately selected weights. 
For simplicity and without loss of generality, all numerical examples in this paper employ NURBS basis functions with unit weights, which are equivalent to B-spline basis functions. 
This choice avoids additional preprocessing associated with exact rational representations and provides a consistent setting for assessing the performance of the proposed formulation.

\subsection{Rolling and unrolling of beams and circular rings with uniform and tapered cross-sections}

\indent \indent 
The first example considers the rolling of a straight tapered beam, as well as the unrolling and re-rolling of an initially open, unstressed circular ring.
Either the straight beam or the ring is subjected to a tip bending moment. 
These two problems are physically opposite to each other.

\begin{figure}[h]
		\centering
		\includegraphics[width=5in]{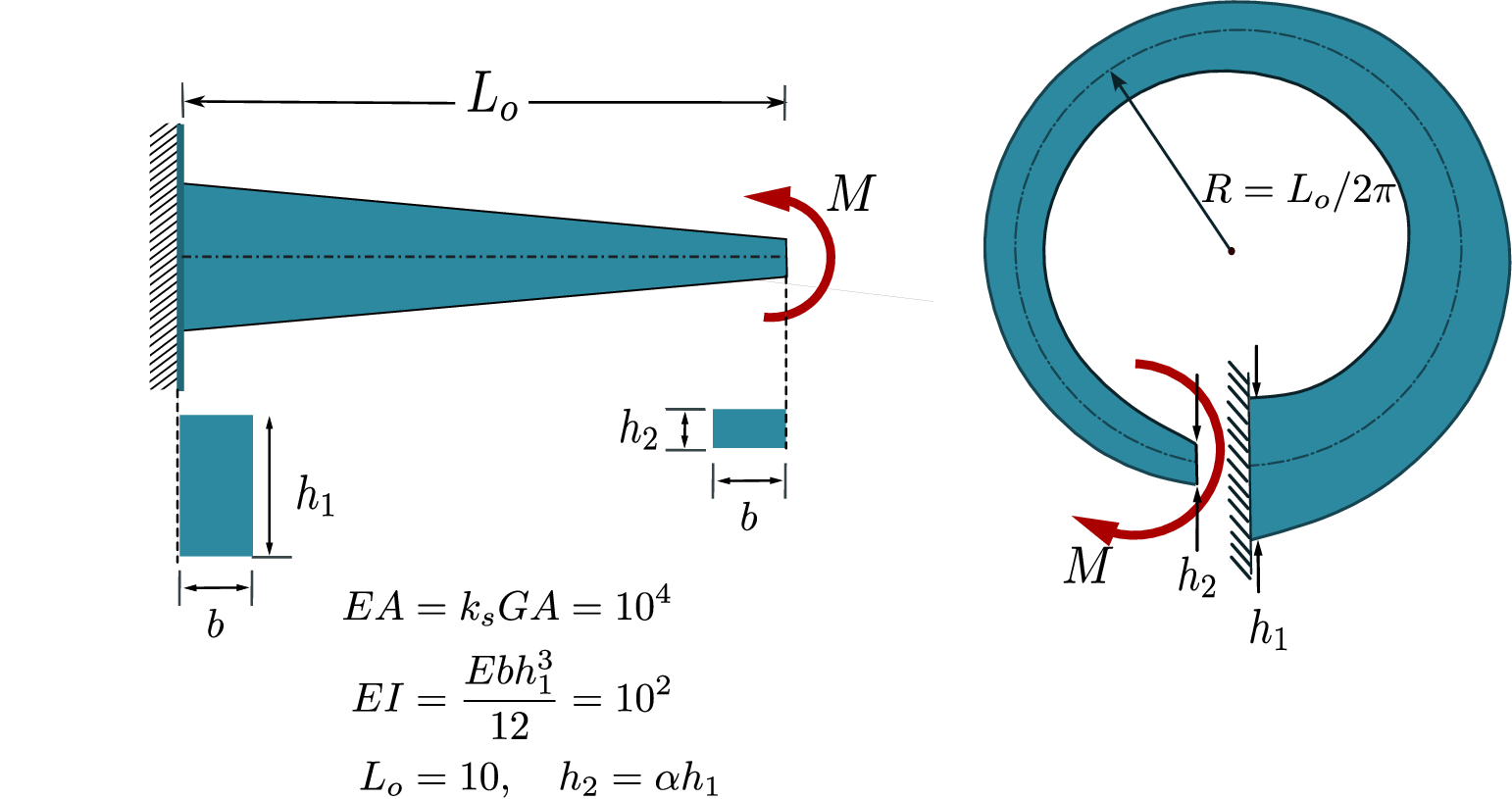}
		\caption{Tapered cantilever beam and circular ring subjected to end bending moments.} 
		\label{fig:rollUnrollSketch}
	\end{figure}

The beam and ring are assumed to have a thickness that varies linearly along the arc-length direction, as illustrated in Fig.\ \ref{fig:rollUnrollSketch}.
The thickness is $h_1$ at the fixed end, and $h_2 = \alpha h_1$ at the tip.
The width of the beam is $b$.
The analytical solution of both rolling and unrolling cases can be obtained from Euler-Bernoulli theory, and the governing equation is given as,
    \begin{equation}
        EI \left[  1 - (1-\alpha) \frac{s_o}{L_o}
        \right]^3 
        \frac{d \theta}{ ds_o} 
        = M, \qquad 0 \le s_0 \le L_o
    \end{equation}
with $I=b{h_1}^3/12$, and boundary condition is $\theta(s_o=0) = 0$. $R=L_o/2\pi$ is the radius of the circular ring. 
The normalized bending moment is defined as $\bar{M} = {M L_o}/{EI}$.
The corresponding analytical solutions are provided in Appendix B.

\begin{figure}[htp]
		\centering
		\subfloat[]{
			\includegraphics[width=3in]{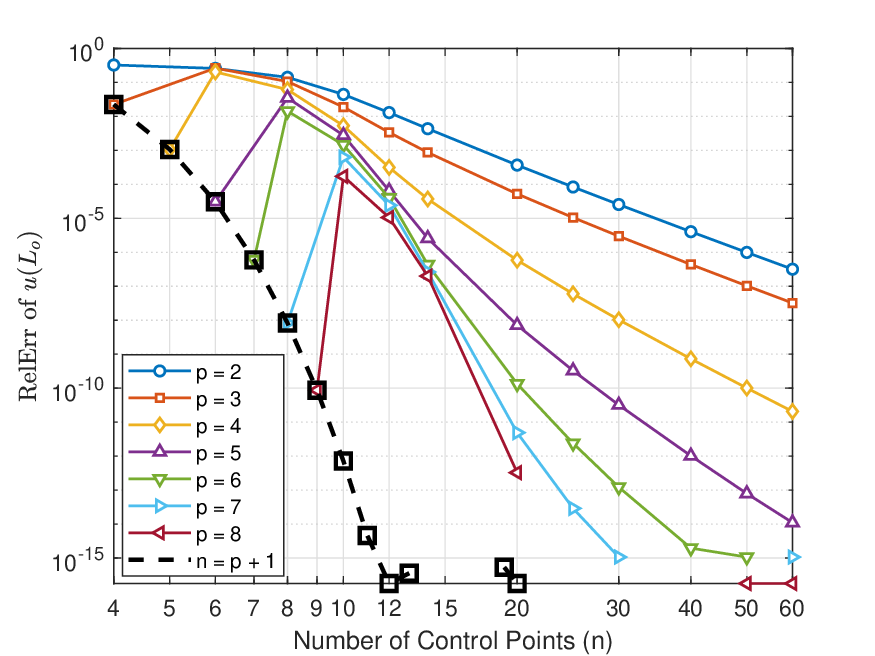}
            \label{fig:converge_uniform_rolling_u}
		}
		\subfloat[]{
			\hspace{-0.2in}
			\includegraphics[width=3in]{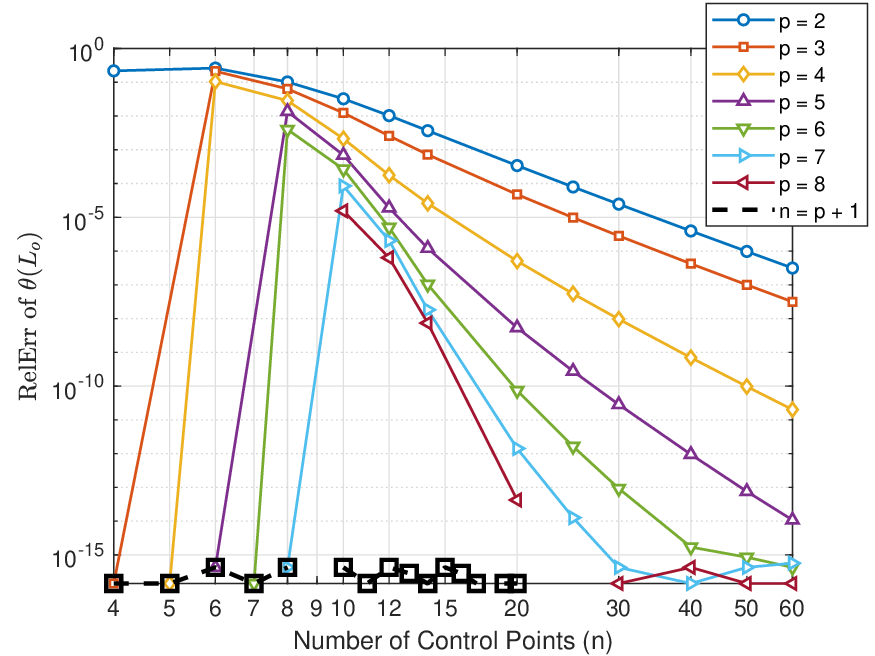}
            \label{fig:converge_uniform_rolling_theta}
		}\\
        \subfloat[]{
			\includegraphics[width=3in]{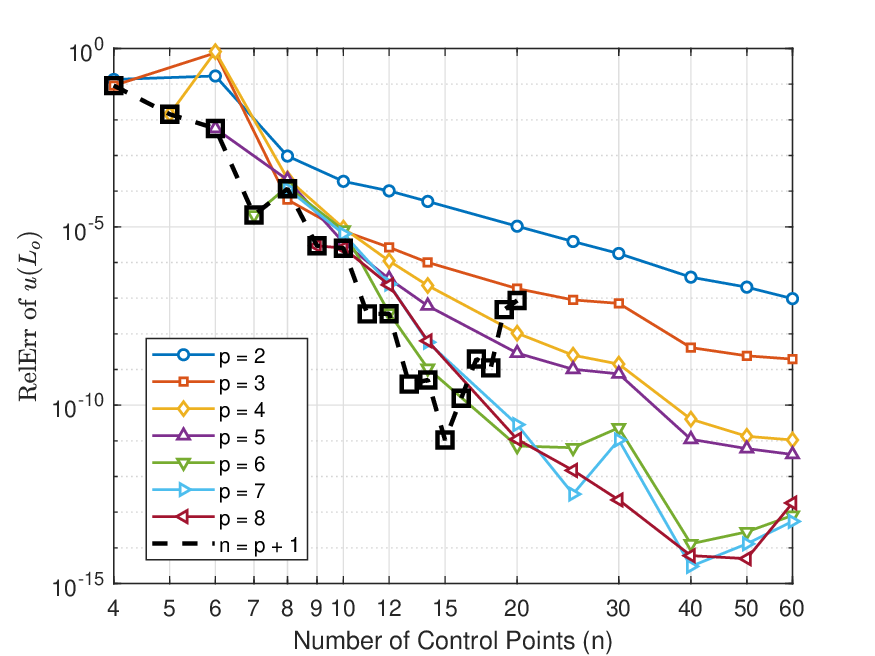}
            \label{fig:converge_uniform_unrolling_u}
		}
		\subfloat[]{
			\hspace{-0.2in}
			\includegraphics[width=3in]{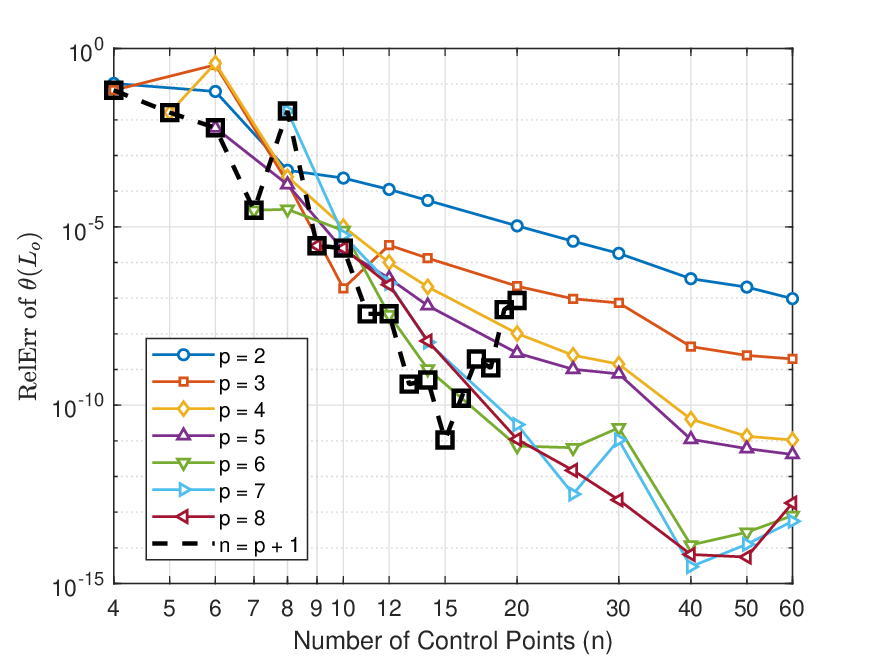}
            \label{fig:converge_uniform_unrolling_theta}
		}\\
		
		\caption{Convergence study of a uniform cantilever beam/ring subjected to an end bending moment using one patch: (a) relative error of tip axial displacement  (b) relative error of tip rotation when rolling a uniform cantilever beam into a circle ($\bar{M} = 2\pi$); (c) relative error of tip axial displacement and (d) relative error of tip rotation when unrolling a uniform ring into a straight beam  ($\bar{M} = -2\pi$). (Note: in (a), RelErr=0 when $p=6:n=60$; $p=7:n=40,50$; $p=8:n=25,30,40$; $p=13:m=14$; $p=14:n=15$; $p=15:n=16$; $p=16:n=17$; $p=17:n=18$. in (b), RelErr=0 when $p=8,n=9$; $p=8,n=25$; $p=17,n=18$).}
		\label{fig:converge_uniform}
	\end{figure}

To describe the initial geometry by using NURBS curves, a set of 100 uniformly distributed data points is first sampled from the initial configuration.
The corresponding control point values for the initial configuration are then determined following the procedure described in Section \ref{sec:decribeInitialShape}.
Fig.\ \ref{fig:converge_uniform} presents the convergence study for two cases: (i) rolling of a straight uniform beam subjected tip concentrated normalized moment $\bar{M} = 2\pi$, and (ii) unrolling of a circular ring with a uniform cross-section subjected to a negative normalized moment $\bar{M}=-2\pi$.
The convergence behavior is assessed using the relative errors of the tip transverse displacement $u(L_o)$ and tip rotation $\theta(L_o)$.

For a quantity $\phi$, the relative error is defined as
\begin{equation}
    \text{RelErr of } \phi = \left| \frac{ \phi_{\text{numer}} - \phi_{\text{analy}}}{\phi_{\text{analy}}} \right|.
\end{equation}
Under the applied moments, the analytical solutions indicate that the straight beam rolls into a circle, while the circular ring unrolls into a straight beam.
In these plots, various combinations of order $p$ and number of control points $n$ are considered. 
Solid lines represent results obtained by fixing the order $p$ while varying the number of control points. 
Dashed lines correspond to the case that number of control points equals the order plus one, namely, $n=p+1$.
Some points are not visible in the plots because the numerical solutions match the analytical solutions up to machine precision (16-th digit), resulting in zero relative error.

Figures.\ \ref{fig:converge_uniform_rolling_u} and \ref{fig:converge_uniform_rolling_theta} 
show the relative errors of the tip transverse displacement and the rotation angle, respectively, for the rolling of a straight beam.
Since the beam is initially straight, the initial geometry is exactly represented.
Therefore, the convergence results presented in these plots reflect the effects of numerical errors, without any geometric discretization error.
For a given basis function order, the mesh with $n=p+1$ exhibits superior accuracy compared to meshes with more control points up to a certain number.
When the number of control points are greater than $p+1$, the accuracy improves monotonically as $n$ increases.
For the mesh with $n=p+1$, the results on $u(L_o)$ show an exponential convergence rate, similar to what is observed in $p$-refinement in conventional finite element methods and in the weak-form quadrature element method \cite{yuan_co-rotational_2019}.
In particular, for the problem considered here, the relative error of the rotation angle
$\theta(L_o)$ reaches the machine precision ($10^{-15} - 10^{-16}$) for all cases with $n=p+1$.

Figures.\ \ref{fig:converge_uniform_unrolling_u} and \ref{fig:converge_uniform_unrolling_theta} show convergence studies for the unrolling of a uniform circular ring into a straight configuration.
Since the initial geometry is circular and cannot be exactly represented by NURBS basis functions with unit weights, the convergence results reflect the combined effects of geometric discretization error and numerical error.
When geometric discretization error is present, the choice $n=p+1$ might not be optimal.
For a given basis function order $p$, the accuracy improves as the number of control points increases. 
For a fixed number of control points, increasing the order $p$ also enhances the accuracy.
However, the improvement becomes less significant for orders higher than $p=7$.
The unusual convergence behavior observed for the $n=p+1$ curve beyond $n=15$ is consistent with the behavior of geometric discretization error when higher-order basis functions are employed.

\begin{figure}[h]
    \centering
    \includegraphics[width=5in]{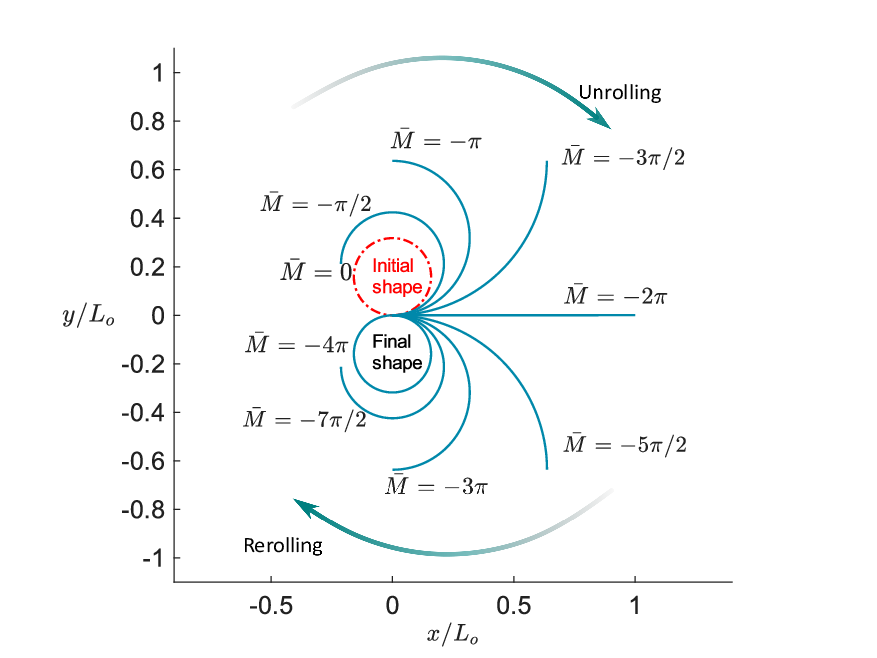}
    \caption{Unrolling and re-rolling a uniform circular cantilever beam using one patch with $p=7, n=8$.}
    \label{fig:Unrolling_and_re-rolling}
\end{figure}

The deformed shapes of a uniform ring subjected to a tip bending moment during the unrolling and rerolling process are shown in Fig.\ \ref{fig:Unrolling_and_re-rolling}.
In Fig.\ \ref{fig:Unrolling_and_re-rolling}, the ring is meshed using a single patch represented by a septic NURBS curve with eight control points. 
For the uniform ring, when subjected to a normalized bending moment $\bar{M} = -2\pi$, the ring unrolls into a straight beam. With a further increase in the magnitude of the applied moment, the ring continues to deform, and when $\bar{M} = -4\pi$, it forms another circular ring in the opposite direction.

\begin{figure}[htp]
		\centering
		\subfloat[]{
			\includegraphics[width=3in]{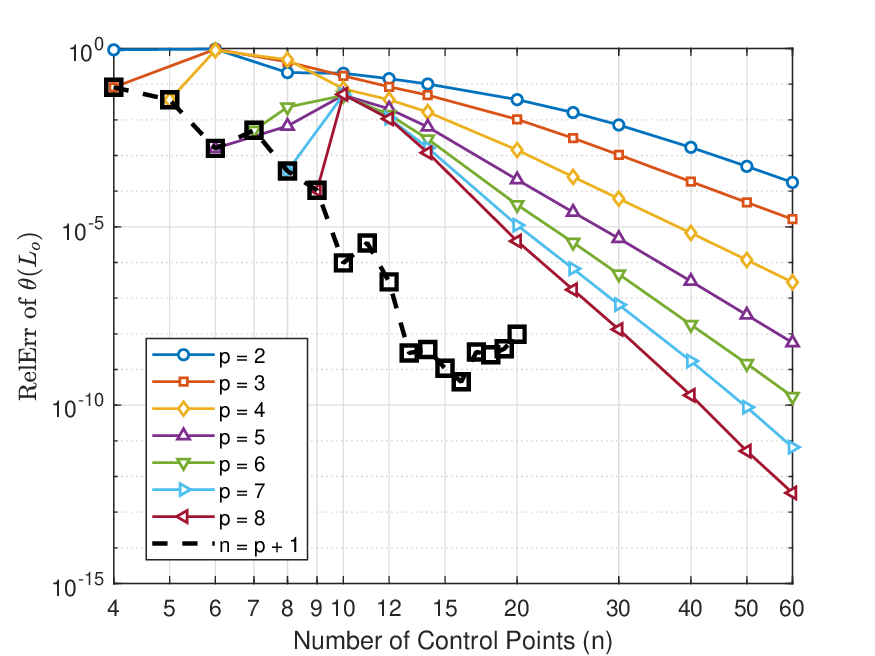}
            \label{fig:converge_a02_unrolling_theta}
		}
		\subfloat[]{
			\hspace{-0.2in}
			\includegraphics[width=3in]{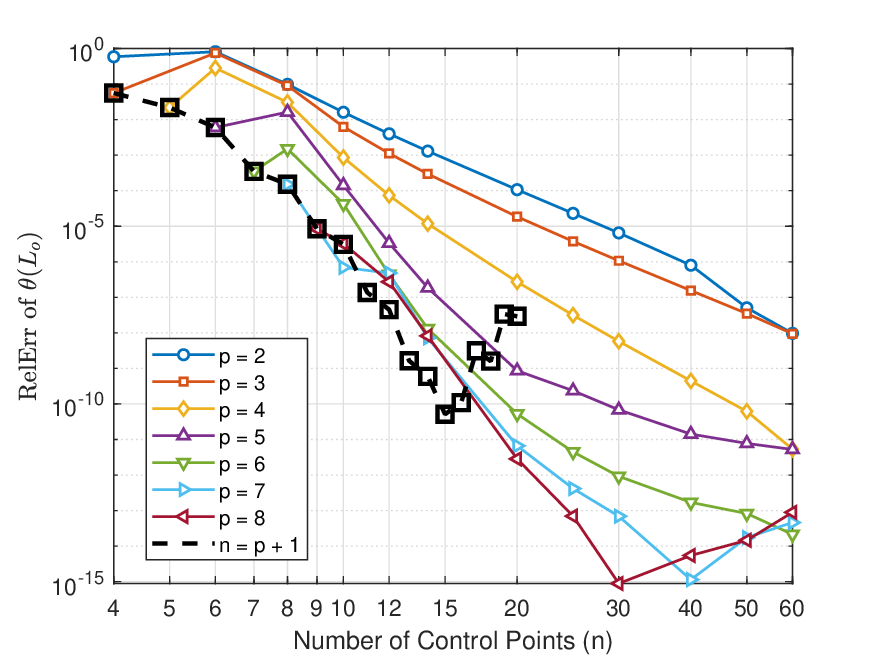}
            \label{fig:converge_a06_unrolling_theta}
		}\\

		\caption{Convergence study of unrolling tapered ring subjected to an end bending moment. (a) $\alpha=0.2$ and $\bar M = {-\pi}/{16}$; (b) $\alpha=0.6$ and $\bar M = -\pi/2$.}
		\label{fig:converge_tapered_unrolling}
	\end{figure}

Furthermore, convergence studies for the unrolling of a tapered ring are shown in Fig.\ \ref{fig:converge_tapered_unrolling}.
A single patch is employed, the same as in the case of unrolling a uniform ring.
The variation in cross-section is accounted for through the values evaluated at the Gauss quadrature points. 
Thus, when comparing Fig.\ \ref{fig:converge_tapered_unrolling} with Figs.\ \ref{fig:converge_uniform_unrolling_u}-\ref{fig:converge_uniform_unrolling_theta}, the same geometric discretization error is retained, since an identical mesh is used.
Fig.\ \ref{fig:converge_tapered_unrolling} further illustrates the effect of higher-order basis functions on numerical accuracy in the presence of a non-uniform cross-section.
It is observed that the presented co-rotational formulation accurately captures the effects of non-uniform cross-sections without requiring special treatment.
Owing to the non-uniformity, higher-order basis functions show more pronounced improvements than those observed in the uniform ring case.
In particular, increasing the order from $p = 6$ to $p = 7$ yields noticeable accuracy gains for the tapered ring, whereas the relative errors for $p = 6$ and $p = 7$ are comparable for the uniform ring.
Similar to the uniform case, for very high orders (i.e., $p > 12$), the accuracy does not improve monotonically and may even deteriorate. 
The non-monotonic behavior of the $n=p+1$ curve in Fig.~\ref{fig:converge_tapered_unrolling} may be associated with the underlying geometric discretization error, since a similar trend is observed in the geometric discretization error when very high-order basis functions are used. 
This behavior may be related to the deterioration of matrix conditioning under very high-order discretizations, which can increase sensitivity to round-off errors and affect numerical robustness \cite{Cottrell2006, Eisentrager2020}.

\subsection{Circular ring subjected to opposite compressive forces}

\begin{figure}[h]
		\centering
		\includegraphics[width=3in]{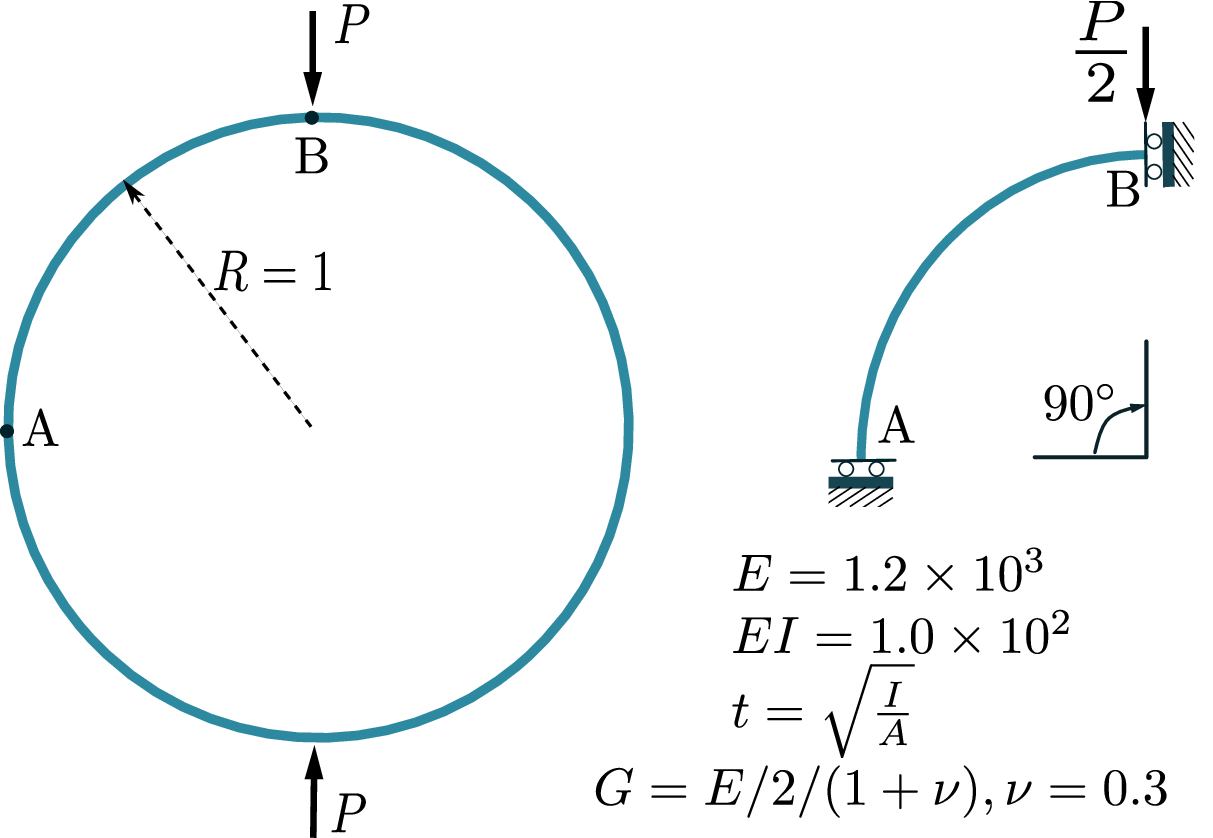}
		\caption{Schematic of a circular ring subjected to compressive forces and the corresponding numerical model exploiting symmetry conditions.}
		\label{fig:quadrant beam subjected to shear force}

	\end{figure}

\indent \indent
We further study the locking behavior of the present formulation by considering a circular ring subjected to two equal compressive concentrated forces, as shown in Fig. \ref{fig:quadrant beam subjected to shear force}.
This benchmark problem is commonly used to assess curved beam elements \cite{borkovic_rotation-free_2018, Armero2024}
Because of symmetry, only a quarter of the ring is modeled.
The circular ring has a unit radius $R=1$, Young's modulus $E=1.2\times 10^3$, and bending stiffness $EI = 1\times 10^2$. 
To investigate locking effects in the thin-rod limit, three value of the cross-sectional area $A$ are considered. 
These rods are characterized by the radius of gyration, $t = \sqrt{I/A}$, which provides a measure of the rod's thickness. 
Three slenderness ratios are considered: $R/t = 10, 100$, and $1000$.
A load of $P = 800$ is applied in all cases.

The deformation obtained using four sets of meshes is presented in Fig.\ \ref{fig:ring_compression_Rt10} - \ref{fig:ring_compression_Rt1000}. 
Two sets employs lower-order $p=2$ NURBS basis functions, while the other two use higher-order $p=7$ basis functions. 
In this example, since $EI$ is fixed and $EA = EI/t^2$, a larger $t$ corresponds to smaller axial and shear stiffnesses, $EA$ and $GA$, respectively, leading to larger deformations.
For the relatively thick case ($R/t = 10$), all meshes yield nearly identical results.
As the slenderness ratio increases, locking effects become evident in meshes using lower-order basis functions, particularly when the number of control points is insufficient.
For $R/t = 1000$, even with 20 control points and $p = 2$, noticeable discrepancies remain compared to the results obtained with higher-order basis functions using fewer control points.
In contrast, the locking effect is negligible when using $p = 7$ basis functions, even with as few as $8$ control points.

The total strain energy with various meshes are shown in Fig.\ \ref{fig:ring_compression_Rt10_energy} - \ref{fig:ring_compression_Rt1000_energy}.
Results computed using ABAQUS with 1000 B21 elements are also included for comparison.
The energy plots lead to the same conclusions as the deformed shapes.
Lower-order basis functions tend to suffer from locking issues as the slenderness ratio increases, while the higher-order basis functions remain free of the locking issue, even with a relatively small number of DOFs in the model.
Interestingly, the total strain energy obtained by ($p=5,n=6$) is closer to the converged values than that obtained with ($p=5,n=8$). 
This observation is consistent with the behavior of meshes using $n = p + 1$ in the previous rolling and unrolling examples.
These results indicate that, although the present formulation is not inherently free of locking, as no special treatment is introduced in the formulation, the locking behavior can be effectively mitigated by employing higher-order NURBS basis functions.
Combined with the convergence studies presented for the beam/ring rolling/unrolling problems, a polynomial order in the range $p = 4\text{--}7$ appears to provide a good balance between accuracy and computational efficiency.

\subsection{Pinned-fixed deep arch subjected to a concentrated force}

\begin{figure}
		\centering
		\includegraphics[width=3.7in]{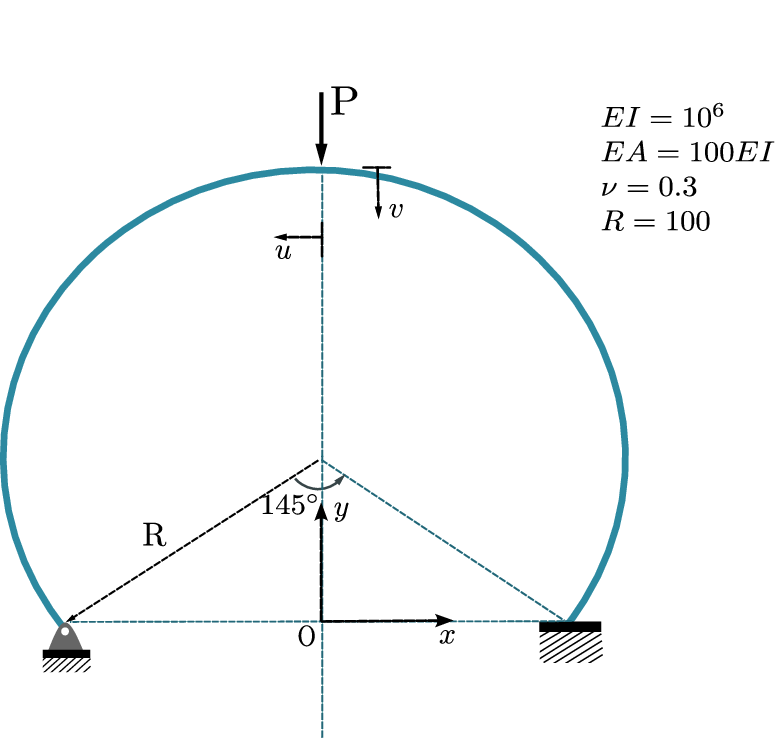}
		\caption{Pinned-fixed deep arch subjected to a concentrated force
        }
		\label{fig:Deep}
	\end{figure}

\indent \indent
The third example investigates the sideways stability behavior of a deep arch.
A deep elastic arch, with one end clamped and the other hinged, is subjected to a concentrated load applied at the crown, as shown in Fig.\ \ref{fig:Deep}. 
This problem is another standard benchmark for geometric nonlinear analysis of curved beam elements undergoing large displacement and large rotation \cite{dadeppo_instability_1975, saje_kinematically_1998,zupan_finite-element_2003}. 
This example was first studied by DaDeppo and Schmidt in \cite{dadeppo_instability_1975}, in which the beam is considered axial inextensible. 

To further examine the performance of the present approach with multiple patches, two meshes are considered.
The first mesh discretizes the entire arch using a single patch with a septic NURBS curve and 60 control points ($p = 7$, $n = 60$).
The second mesh uses two patches, each represented by a septic  NURBS curve with 30 control points.
The two patches are connected through shared control points, and the rigid connection is automatically enforced via shared degrees of freedom.

\begin{figure}
		\centering
		\includegraphics[width=4.0in]{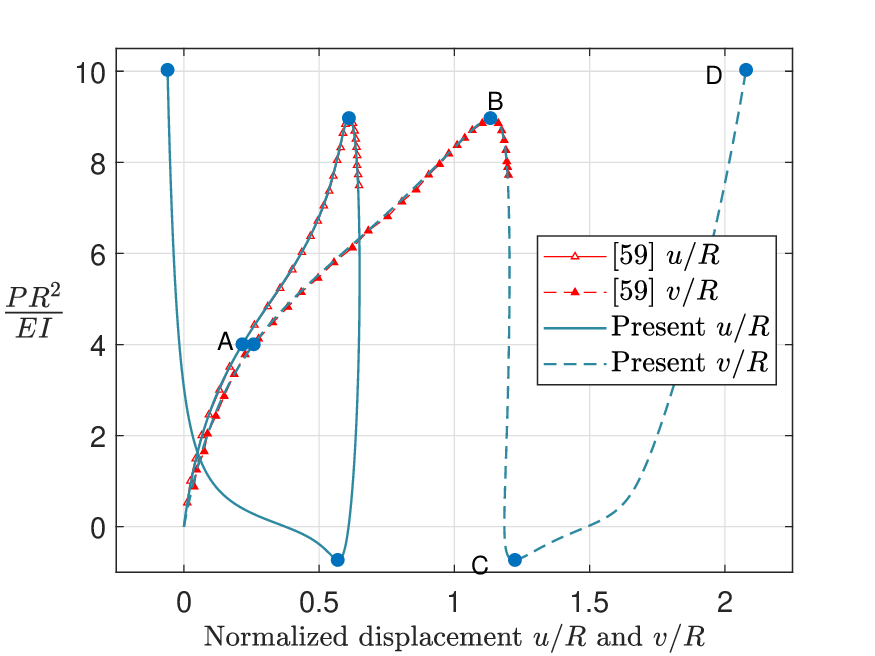}
		\caption{Normalized load-displacement curve of finned-fixed deep arch subjected to a concentrated force with one patch $p=7: n=60$. }
		\label{fig:Deep_ref}
	\end{figure}

\begin{figure}
		\centering
		\includegraphics[width=4in]{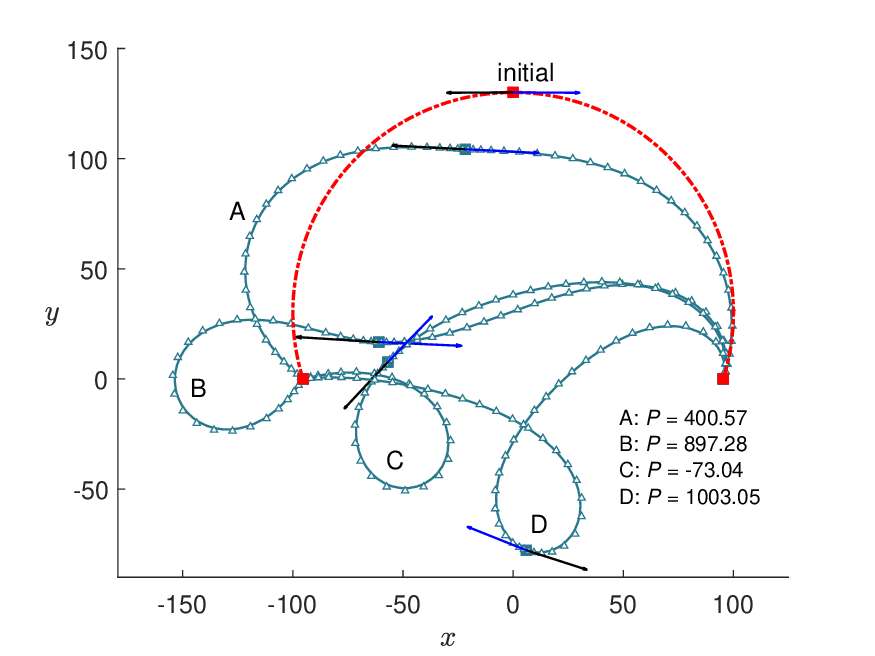}
		\caption{Deformed axes of pinned-fixed deep arch subjected to a concentrated force with two patches, with $p=7$ and $n=30$ for each patch. (Solid squares indicate edge control points, which coincide with the start and end of each patch. Empty triangles denote interior control points.)}
		\label{fig:Deep_shape}
	\end{figure}

The vertical and horizontal displacements at the load application point, against the applied load, are shown in Fig.\ \ref{fig:Deep_ref}.
The two meshes yield identical results. Therefore, only the results obtained with the single-patch model are presented.
For validation, results from the literature \cite{dadeppo_instability_1975} are also included, showing good agreement.
Several deformed configurations are shown in Fig.\ \ref{fig:Deep_shape}. 
Configurations A, B, C, and D correspond to the points marked in Fig.\ \ref{fig:Deep_ref}.
Fig.\ \ref{fig:Deep_shape} presents the deformation obtained using the two-patch discretization, which is identical to that obtained with the one-patch mesh.
It is observed that the rigid connection at the arch midpoint is automatically enforced through the shared control points. 
Both single- and two-patch discretizations accurately capture the complex deformation, including folded shapes.
Configurations B and C correspond to limit points, and the associated loads are listed in Table \ref{tab:deep}.
The limit load values obtained using both meshes (single- and two-patch) are nearly identical.
The limit load at point B, which is commonly reported in the literature, agrees well with the reference results.

\subsection{Free fall of curved pendulums}

\begin{figure}
		\centering
		\includegraphics[width=3.in]{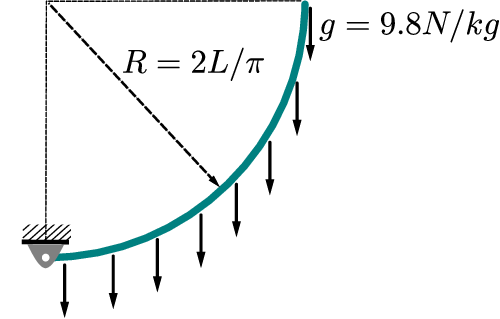}
		\caption{Curved beam pendulum. }
		\label{fig:curved_beam_pendulum}
	\end{figure}

\indent \indent 
In this example, the dynamic response of freely falling curved beam pendulums, as shown in Fig.~\ref{fig:curved_beam_pendulum}, is considered.
The pendulums are subjected solely to gravity.
Similar problems have been studied in the literature \cite{sugiyama_gradient_2010,pan_geometric_2012,patel_locking_2018,Li2023,Wang2025,zheng_series_2025}.
The geometry and material parameters are adopted from Sugiyama et al. \cite{sugiyama_gradient_2010}.

The beam has a length of $L = 1$ m and a radius of curvature $R = 2L/\pi$.
The cross-sectional area $A = 2.514 \times 10^{-4}~\text{m}^2$.
The second moment of area is $I=2.514\times10^{-8}~\text{m}^4$, and the rotary inertia is taken as $I_{\theta}=I$.
The material density is $\rho = 7200~\text{kg/m}^3$ and gravity is $g = 9.8$ N/kg.
Two pendulums with different stiffnesses are considered: one stiff curved beam with Young's modulus $E_{\text{stiff}} = 100$ GPa and one flexible curved beam with $E_{\text{flex}} = E_{\text{stiff}} / 10^4 =10$ MPa. 
The Poisson's ratio is $\nu = 0.3$ and the shear modulus is given by $G = E / [2(1+\nu)]$ for both cases.

\begin{figure}[htp]
    \centering

    \subfloat[]{
        \includegraphics[width=4.5 in]{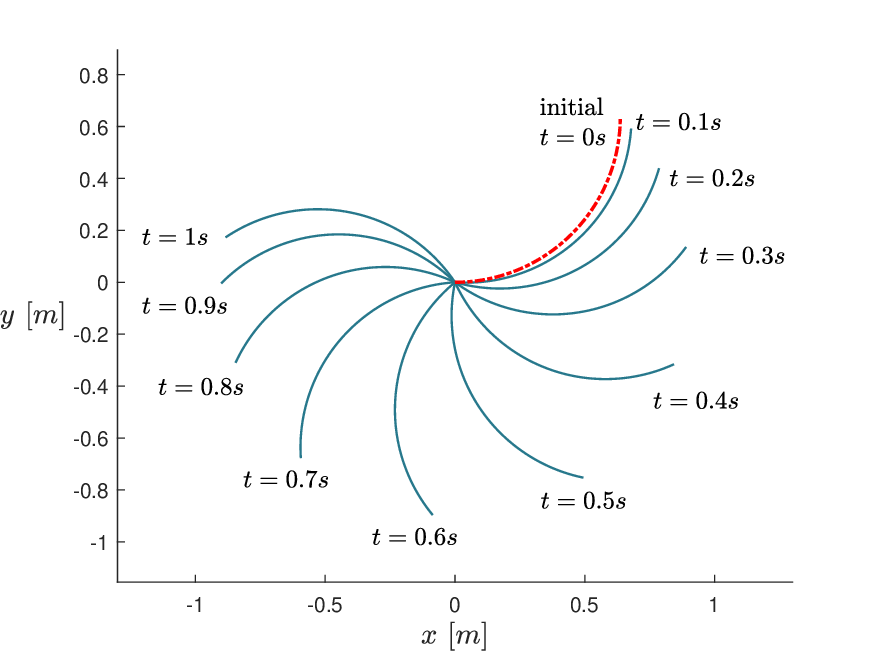}
        \label{fig:stiffer_curved_beam_pendulum_deformed_shape}
    }\\
    \subfloat[]{
        \includegraphics[width=4.5 in]{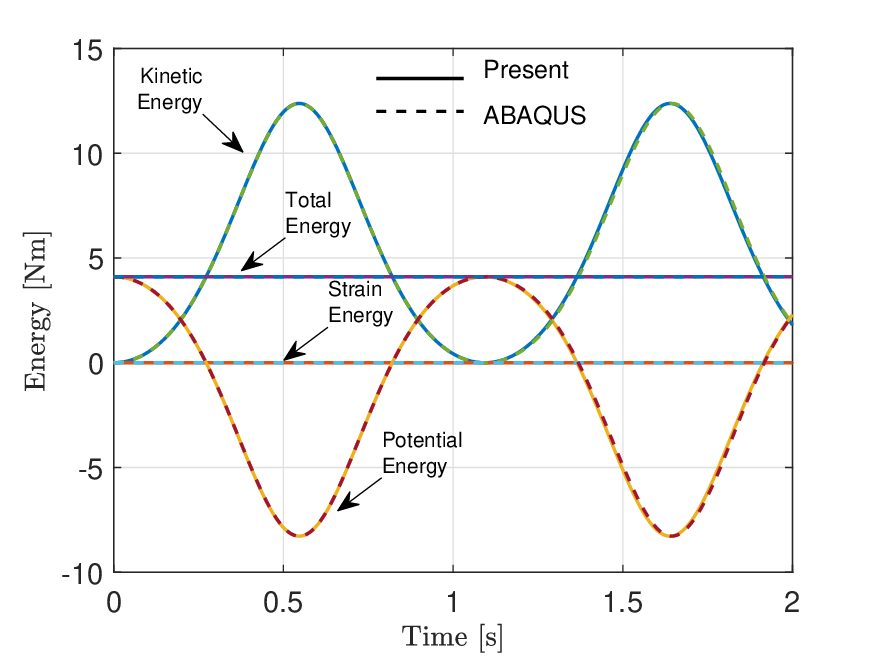}
        \label{fig:stiffer_curved_beam_pendulum_energy}
    }

    \caption{Dynamic response of a stiff curved beam pendulum using one patch with $p=7$ and $n=8$: 
    (a) deformed configuration at selected times; 
    (b) time histories of the energies.}

    \label{fig:stiff_beam_pendulum_deformed_shape_and_v}
\end{figure}

The dynamic response of the stiff curved pendulum is shown in Fig.\ \ref{fig:stiff_beam_pendulum_deformed_shape_and_v}. 
As illustrated in Fig.\ \ref{fig:stiffer_curved_beam_pendulum_deformed_shape}, the beam is sufficiently stiff that its motion closely resembles rigid body rotation about the pinned end.
Thus, a single patch using a septic NURBS curve with eight control points is sufficient to describe both the initial and deformed configurations.
The energy evolution is shown in Fig. \ref{fig:stiffer_curved_beam_pendulum_energy}, where the zero reference for gravitational potential energy is taken at the plane $y = 0$ m.
The center of mass of the beam is initially located at $y=R(1-2/\pi)$.
For validation, the same pendulum is analyzed with ABAQUS/Standard (dynamic implicit) with 200 uniform B21 beam elements.
The energy evolution obtained from the present approach agree well with those from by ABAQUS. 
Since the motion is dominated by rigid body rotation, the strain energy remains nearly zero, and the total energy is exchanged primarily between gravitational potential energy and kinetic energy.
The motion of the stiff pendulum exhibits approximately harmonic motion.

\begin{figure}[htp]
    \centering

    \subfloat[]{
        \includegraphics[width=4.5 in]{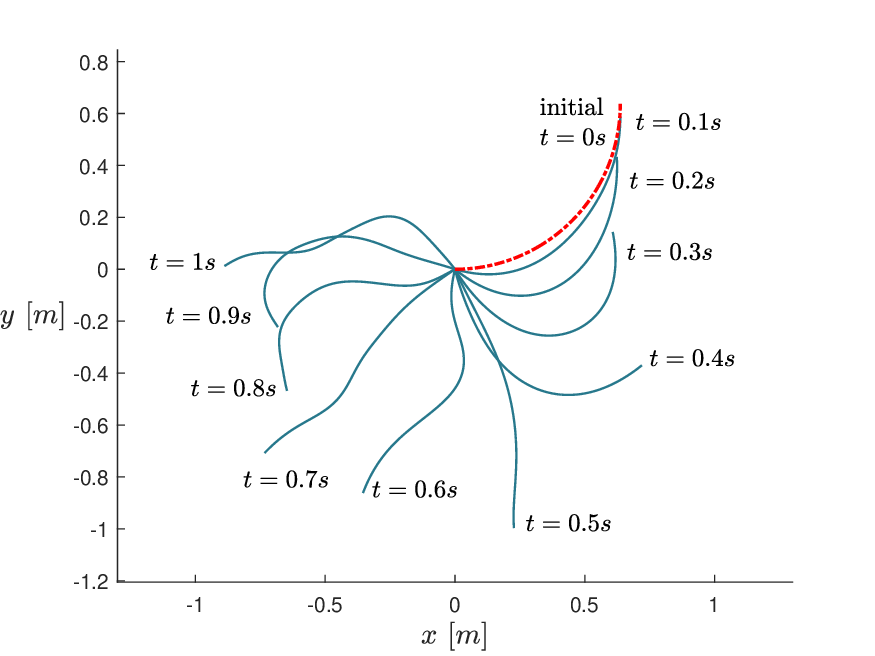}
        \label{fig:softer_curved_beam_pendulum_deformed_shape}
    }\\
    \subfloat[]{
        \includegraphics[width=4.5 in]{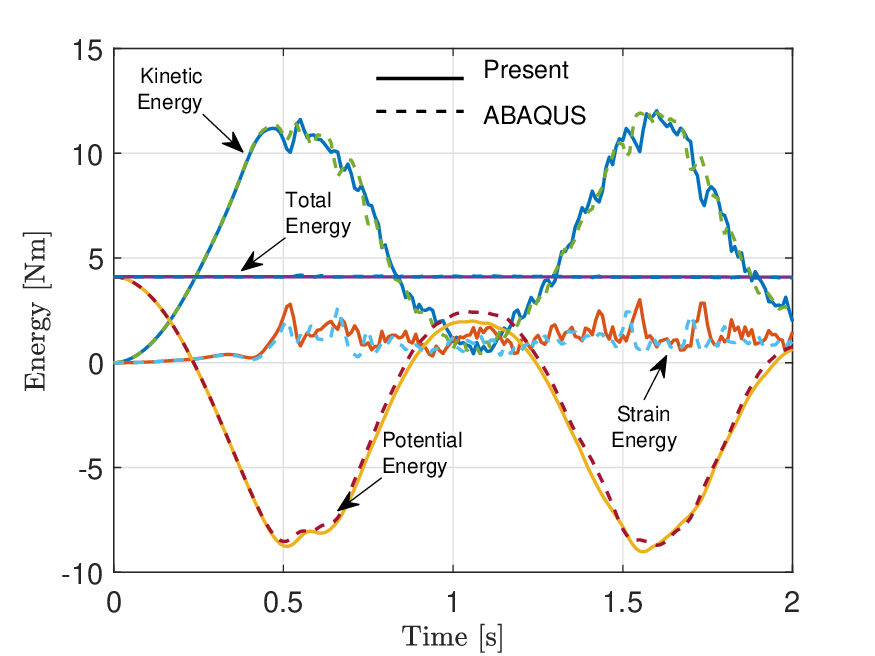}
        \label{fig:softer_curved_beam_pendulum_energy}
    }

    \caption{Dynamic response of a flexible curved beam pendulum using three patches with $p=7$ and $n=8$ for each patch: 
    (a) deformed configuration at selected times; 
    (b) time histories of the energies.}
    \label{fig:soft_beam_pendulum_deformed_shape_and_v}
\end{figure}

The dynamic response of the flexible curved pendulum is shown in Fig.\ \ref{fig:soft_beam_pendulum_deformed_shape_and_v}.
In contrast to the stiff case, a more complex response is observed.
To capture the complex deformed shapes, the flexible pendulum is discretized with three patches.
Each patch uses septic NURBS basis functions and eight control points ($p = 7,n=8$), resulting in a total of 22 control points and 66 DOFs.
At the early stage of motion (around $t<0.3$ s), the midspan of the beam exhibits larger displacements than the tip due to its flexibility, whereas in the stiff case, the tip moves more due to rigid body rotation.
Significant elastic deformation develops over the time, and the response becomes a combination of rigid body motion and elastic deformation.
During the free fall of the flexible pendulum, the strain energy is no longer negligible, and rapid oscillations are observed in its time history.
The total energy is exchanged among gravitational potential energy, strain energy, and kinetic energy. 
Due to the flexibility of the beam, both the motion and the energy components exhibit oscillatory behavior superimposed on the overall harmonic response.

\subsection{Dynamic relaxation of a pre-wound spiral beam}

\indent \indent The final example investigates the dynamic relaxation behavior of a spiral beam, in which the curvature is non-uniform.
A similar problem involving a spiral beam subjected to a static bending moment was studied by Vo et al. \cite{vo_total_2020}.

The geometry of the spiral beam is defined by
\begin{equation}
		\begin{array}{ll}
			& x(\varphi  ) = R_0(1 + \varphi )\cos \varphi ,\\
			& y(\varphi ) =  R_0(1 + \varphi )\sin \varphi , \qquad \varphi \in \left[ {0,6\pi } \right].
		\end{array}
\end{equation}
The initial radius $R_0=5$ mm. 
The spiral beam is made of steel with Young's modulus $E = 200$ GPa, Poisson's ratio $\nu = 0.3$, and density $\rho = 7850\,\text{kg/m}^3$.
The cross-sectional thickness and width are $h = 2$ mm and $b = 15$ mm, respectively.
The beam is pre-loaded with a torsional moment $M=1.5$ Nm.
The applied moment is then removed, and the beam undergoes free dynamic relaxation and shows torsional oscillations.
Thus, the spiral beam behaves as like a torsional spring.
To accurately represent the complex geometry, the entire spiral beam is discretized using ten patches, each modeled with septic-order NURBS basis functions and eight control points.
The pre-wound configuration is first obtained by solving a static problem using the Newton--Raphson method.
This configuration is then taken as the initial condition for the subsequent dynamic analysis.
The dynamic response is computed using the Newmark method, with a fix time step size $\Delta t=2.5\times 10^{-4}$ s. 

\begin{figure}[htp]
		\centering
		\subfloat[]{
			\includegraphics[width=5in]{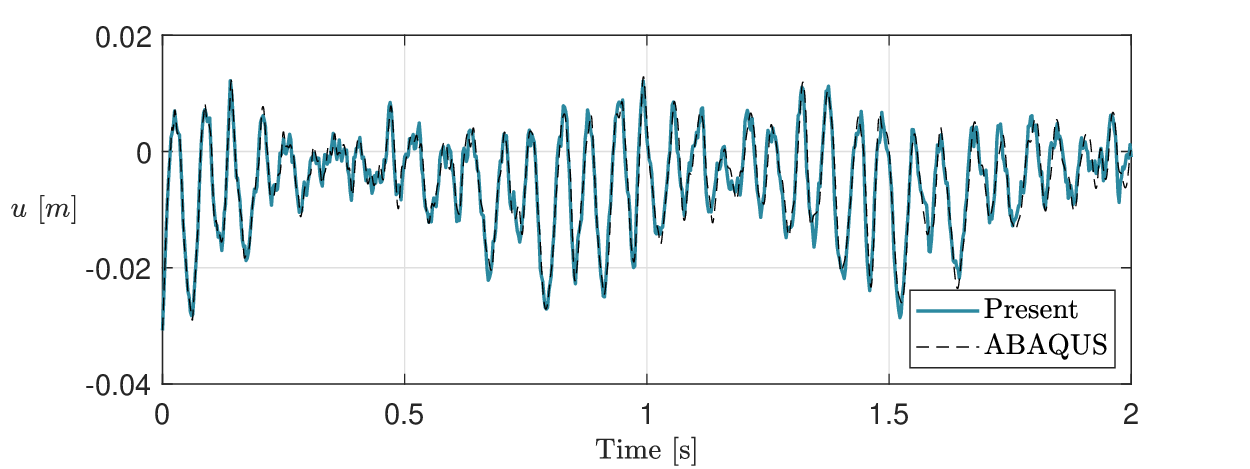}
            \label{fig:Spiral_2s_u_displ_vs_time}
		}\\
		\subfloat[]{
			\includegraphics[width=5in]{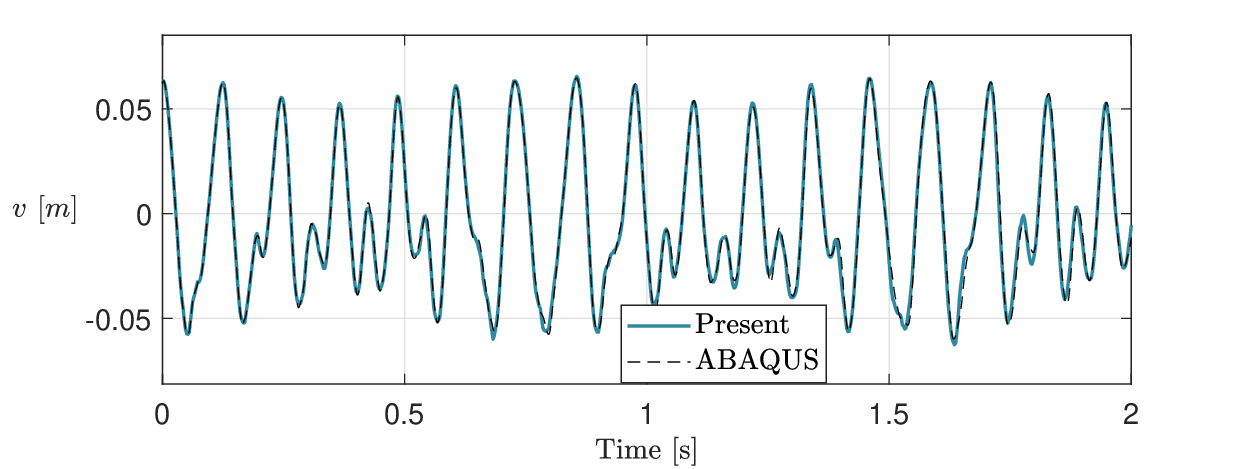}
            \label{fig:Spiral_2s_v_displ_vs_time}
		}\\
        \subfloat[]{
			\includegraphics[width=5in]{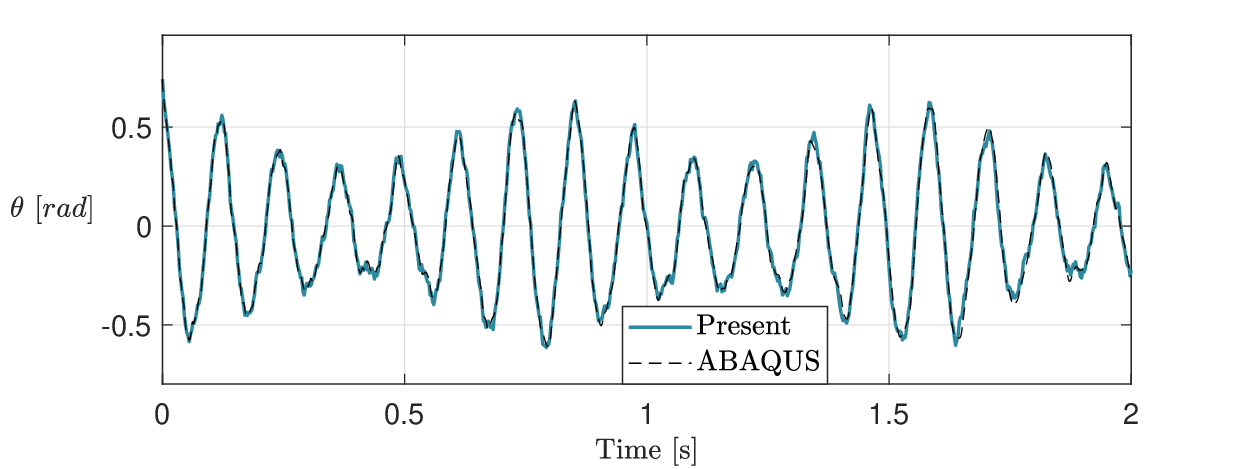}
            \label{fig:Spiral_2s_theta_displ_vs_time}
		}
    \caption{The first $2s$ of tip displacement of spiral spring modeling using 10 patches ($p=7$ and $n=8$ for each patch): (a) horizontal displacement $u$, (b) vertical displacement $v$, (c) rotational angle $\theta$. 
    }
		\label{fig:Spiral_2s_displ_vs_time}
        
\end{figure}

\begin{figure}[htp]
\centering

\subfloat[]{
\includegraphics[width=2.6in]{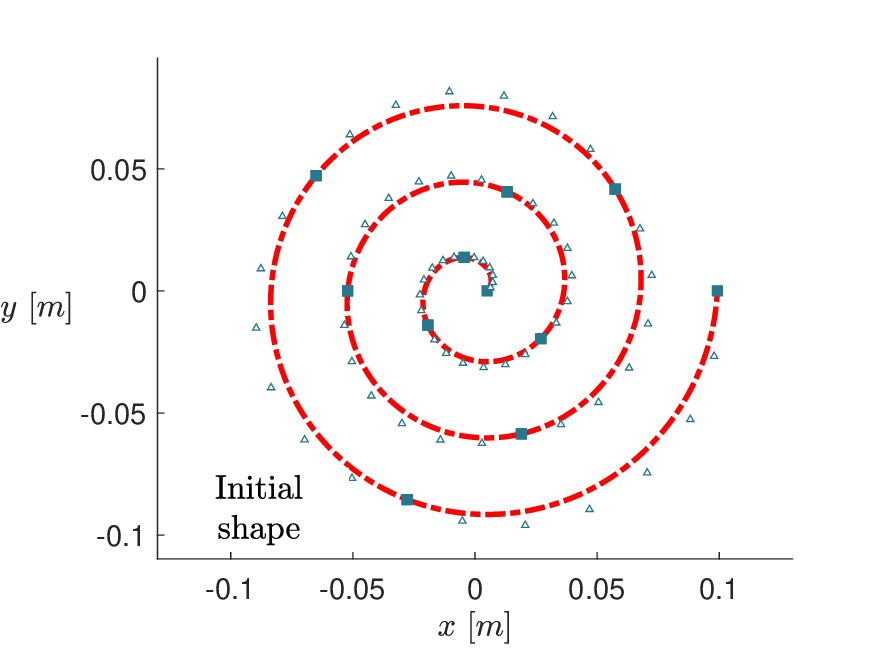}
\label{fig:Spiral_initial_shape}
}
\subfloat[]{
\includegraphics[width=2.6in]{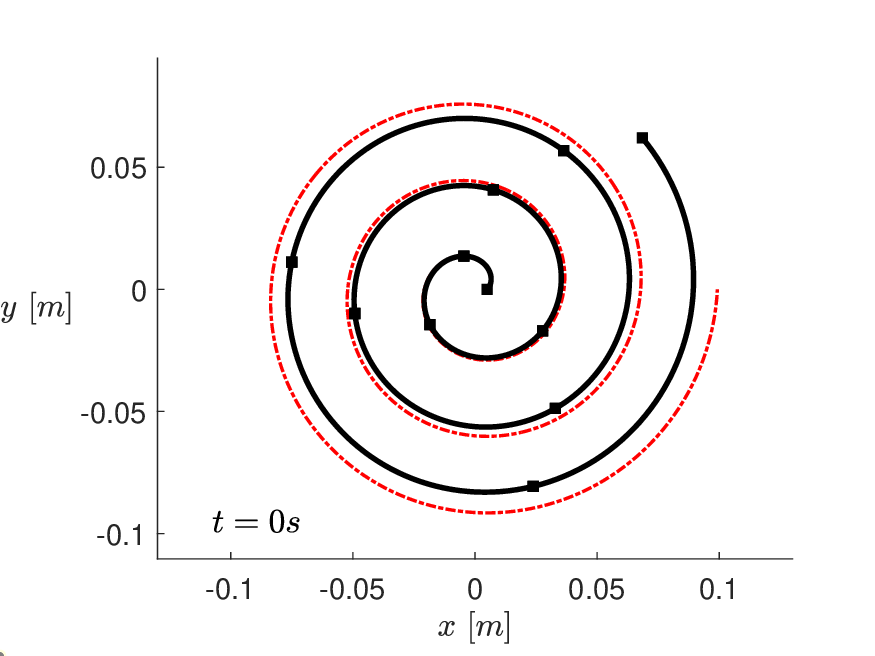}
\label{fig:Spiral_preload_M_static_deform}
}\\[-2mm]

\subfloat[]{
\includegraphics[width=2.6in]{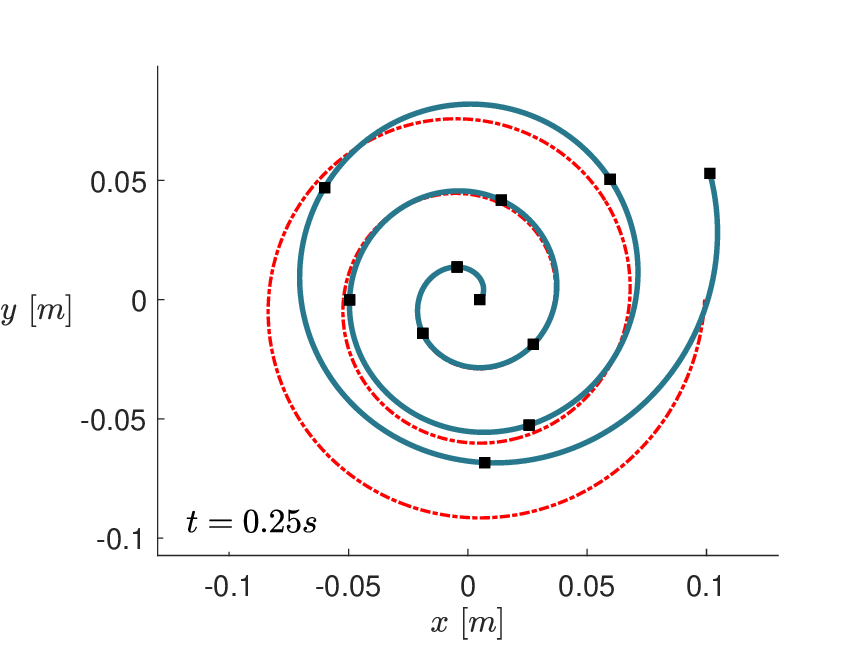}
\label{fig:Spiral_dynamic_dot25s}
}
\subfloat[]{
\includegraphics[width=2.6in]{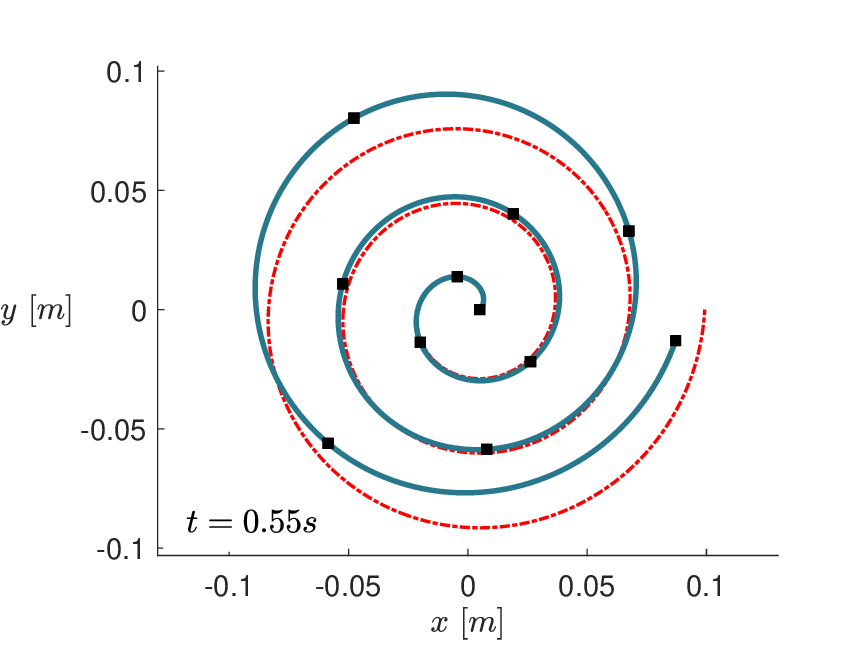}
\label{fig:Spiral_dynamic_dot55s}
}\\[-2mm]

\subfloat[]{
\includegraphics[width=2.6in]{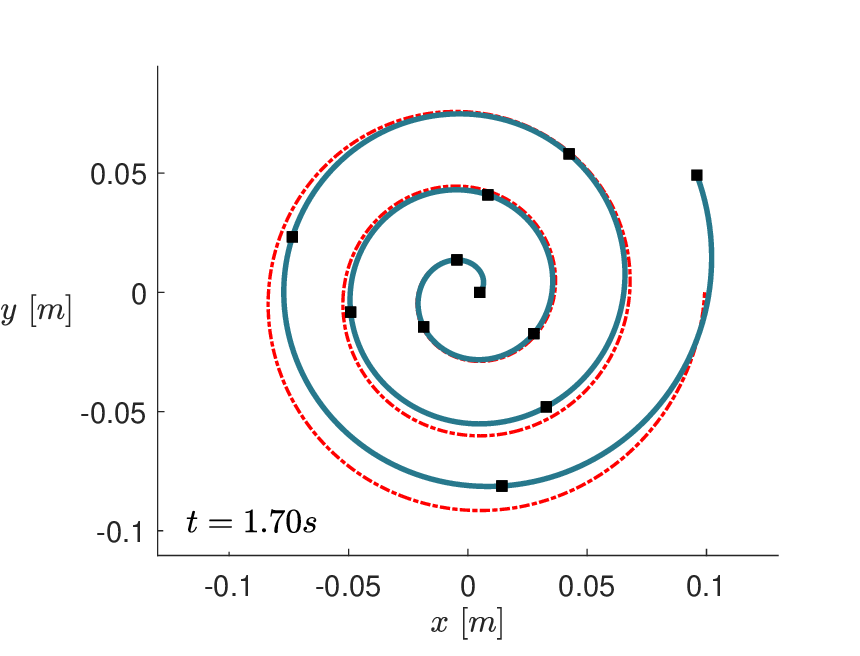}
\label{fig:Spiral_dynamic_1dot7s}
}
\subfloat[]{
\includegraphics[width=2.6in]{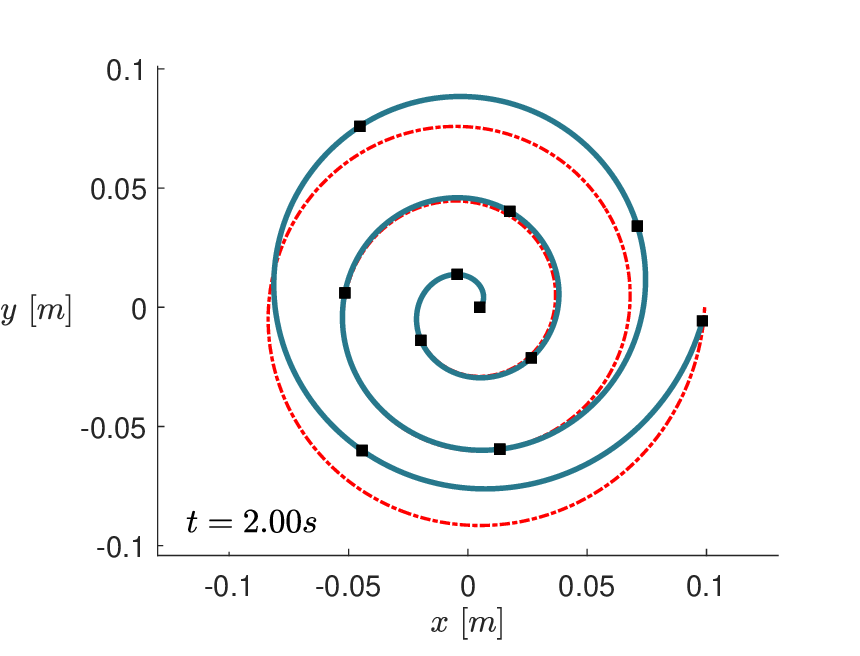}
\label{fig:Spiral_dynamic_2sm}
}

\caption{Deformed shapes of the spiral spring modeling using 10 patches ($p=7$ and $n=8$ for each patch): (a) initial shape with location of control points; (b) preloaded configuration under a torsional moment $M = 1.5\,\mathrm{Nm}$; (c-f) deformed shapes at $t = 0.25\,\mathrm{s}$, $0.55\,\mathrm{s}$, $1.7\,\mathrm{s}$ and $2\,\mathrm{s}$.
The solid square markers are the edge control points of each patch.
}
\label{fig:Spiral_10s_deformed_shape}

\end{figure}

The tip displacement and rotation during the first 2 seconds are shown in Fig.\ \ref{fig:Spiral_2s_displ_vs_time}.
For verification purposes, the results obtained by ABAQUS using 1000 B21 elements are also included. 
It is seen that the tip movement predicted by the present approach agrees well with that obtained from ABAQUS.
The motion of the spiral beam after removing the applied torsing moment shows a combination of multiple vibration modes with different frequencies.
Applying the Fast Fourier Transform (FFT) to the dynamic response shows that the dominant frequency is 8.22 Hz. 
The first five natural frequencies obtained by both the present method and ABAQUS are listed in Table \ref{tab:spiral_frequencies}. 
These results are consistent with the dominant frequency obtained from the dynamic response.
Representative deformed configurations during the pre-winding and relaxation stages are presented in Fig.\ref{fig:Spiral_10s_deformed_shape}. 
It is seen that the present corotational formulation using higher-order basis functions can accurately capture the motion of beams with non-uniform curvature.


\section{Conclusion }
\label{section:5}

\indent \indent
This work presents a novel co-rotational formulation for the analysis of arbitrarily shaped planar beams undergoing large displacements and rotations.
Higher-order basis functions are employed to represent both the initial geometry and the displacement field.
By introducing an auxiliary reference configuration, the formulation utilizes local frames that co-rotate with differential elements across the reference, initial, and current configurations.
This approach ensures that the resulting strain measure is objective and independent of the choice of reference configuration.
While the framework is general, this study adopted NURBS basis functions to leverage their superior geometric representation, smoothness, and ability to efficiently generate higher-order approximations.

Five numerical examples, including static, post-buckling, dynamic, and combined static-dynamic cases, are investigated to assess the convergence behavior, locking performance, and capability of the proposed formulation in capturing large deformation responses of structures with complex initial geometries.
The results demonstrate that the use of higher-order basis functions can effectively mitigate locking without requiring additional treatment.
The proposed formulation performs well for both thick and slender beams.
For open knot vectors, basis function orders in the range $4 \le p \le 7$  are found to provide an optimal balance between accuracy and computational efficiency.
Furthermore, meshes with the number of control points equal to $p+1$ exhibit particularly favorable accuracy, in some cases outperforming discretizations with a larger number of control points.
The presented formulation is shown to accurately capture large rotations and deformations, even for beams with highly non-uniform curvature.

The proposed high-order co-rotational formulation is general and not restricted to a specific choice of basis functions.
Although NURBS are employed in this work, the formulation can be readily extended to other higher-order discretization frameworks.
For instance, a co-rotational arbitrarily shaped weak-form quadrature element can be constructed by implementing Lagrange polynomials with Gauss-Lobatto-Legendre (GLL) nodes within the presented co-rotational framework.
This work thus provides a unified and efficient foundation for the nonlinear analysis of complex, curved beam-type structures, bridging the gap between geometric fidelity and high-order accuracy within the co-rotational framework.

\section*{Acknowledgments}
	
The financial support from the U.S. National Science Foundation under award number CMMI-2502037 is gratefully acknowledged.

	
	\bibliographystyle{unsrt}
	\bibliography{biblio}
	\clearpage
	
	
\bigskip
\section*{Appendix A: Tangent stiffness matrix}
\label{chap:appendixA}

\indent \indent
The tangent stiffness matrix ${\bm K}_T$ is defined in Eq.~(\ref{equ:tangentStiffMatrix}).
The explicit expression of the ${\bm K}_{T,jk}$, associated with the $j$-th and $k$-th control points, is

\begin{subequations}
    \begin{equation}
		\left[ {{{\bm K}_{T,jk}}} \right] =  \left[ {\begin{array}{*{20}{c}}
				{K_{T,jk}^{11}}&{}&{K_{T,jk}^{12}}&{}&{K_{T,jk}^{13}}\\
				{}&{}&{}&{}&{}\\
				{K_{T,jk}^{21}}&{}&{K_{T,jk}^{22}}&{}&{K_{T,jk}^{23}}\\
				{}&{}&{}&{}&{}\\
				{K_{T,jk}^{31}}&{}&{K_{T,jk}^{32}}&{}&{K_{T,jk}^{33}}
		\end{array}} \right]
	\end{equation}
	with
\begingroup
\allowdisplaybreaks
\begin{flalign}
K_{T,jk}^{11} &= \int_0^{L_r}  N'_{j,p} N'_{k,p}  \left[ EA \cos^2\theta_c + \kappa_s GA \sin^2\theta_c \right]
 \frac{ds_r}{ds_o} ds_r, & \\  
K_{T,jk}^{22} &= \int_0^{L_r}  N'_{j,p} N'_{k,p}  \left[ EA \sin^2\theta_c + \kappa_s GA \cos^2\theta_c\right]
\frac{ds_r}{ds_o} ds_r , & \\ 
K_{T,jk}^{33} &= \int_0^{L_r} \left\{
     EI \, N'_{j,p} N'_{k,p} \right.  +  N_{j,p} N_{k,p}  
 \Bigl[
        EA  (R2_c^2 - (\varepsilon_{c} - \varepsilon_{o}) R1_c \bigr)  \nonumber &\\*
       & \hspace{12 em} + \left. \kappa_s GA \bigl( R1_c^2 - (\gamma_{c} - \gamma_{o}) R2_c \bigr)
    \Bigr] \right\}  \frac{ds_r}{ds_o} ds_r, & \\
K_{T,jk}^{12} &= K_{T,jk}^{21} = \int_0^{L_r} 
N'_{j,p} N'_{k,p} \sin\theta_c \cos\theta_c  \left[ EA - \kappa_s GA \right]
 \frac{ds_r}{ds_o} ds_r, & \\ 
K_{T,jk}^{13} &= \int_0^{L_r} N'_{j,p} N_{k,p}  \Bigl\{
    EA \bigl[ R2_c \cos\theta_c - (\varepsilon_{c} - \varepsilon_{o})\sin\theta_c \bigr] \nonumber \\*
&\hspace{8 em} - \kappa_s GA \bigl[ -R1_c \sin\theta_c + (\gamma_{c} - \gamma_{o})\cos\theta_c \bigr]
\Bigr\} \, \frac{ds_r}{ds_o} ds_r , & \\
K_{T,jk}^{31} &= \int_0^{L_r} N'_{k,p} N_{j,p}  \Bigl\{ EA \bigl[ R2_c \cos\theta_c - (\varepsilon_{c} - \varepsilon_{o})\sin\theta_c \bigr] \nonumber \\*
&\hspace{8 em} - \kappa_s GA \bigl[ -R1_c \sin\theta_c + (\gamma_{c} - \gamma_{o})\cos\theta_c \bigr]
\Bigr\} \, \frac{ds_r}{ds_o} ds_r , & \\
K_{T,jk}^{23} &= \int_0^{L_r} N'_{j,p} N_{k,p}  \Bigl\{
    EA \bigl[ R2_c \sin\theta_c + (\varepsilon_{c} - \varepsilon_{o})\cos\theta_c \bigr] \nonumber \\*
&\hspace{8 em} + \kappa_s GA \bigl[ -R1_c \cos\theta_c - (\gamma_{c} - \gamma_{o})\sin\theta_c \bigr]
\Bigr\} \,  \frac{ds_r}{ds_o} ds_r, & \\
K_{T,jk}^{32} &= \int_0^{L_r} N'_{k,p} N_{j,p}  \Bigl\{
    EA \bigl[ R2_c \sin\theta_c + (\varepsilon_{c} - \varepsilon_{o})\cos\theta_c \bigr] \nonumber \\*
&\hspace{8 em} + \kappa_s GA\bigl[ -R1_c \cos\theta_c - (\gamma_{c} - \gamma_{o})\sin\theta_c \bigr]
\Bigr\} \, \frac{ds_r}{ds_o} ds_r . &
\end{flalign}
\end{subequations}

\endgroup

\bigskip
\section*{Appendix B: Analytical solution of rolling and unrolling tapered beam and ring }
\label{chap:appendixB}

\indent \indent
The analytical solution of both rolling and unrolling cases can be obtained from Euler-Bernoulli theory and the governing equation is given as,
    \begin{equation}
        EI \left[  1 - (1-\alpha) \frac{s_o}{L_o}
        \right]^3 
        \frac{d \theta}{ ds_o} 
        = M, \qquad 0 \le s_0 \le L_o
    \end{equation}
with $I=b{h_1}^3/12$, and bounday condition is $\theta(s_o=0) = 0$. $R=L_o/2\pi$ is the radius of the circular ring. 

The analytical solution of the rotation angle is
\begin{equation}
    \theta (s_o) = \left\{
    \begin{array}{lc}
         \frac{M}{EI}  s_o, \qquad \alpha = 1   \\
          \frac{L_o^3}{2(1 - \alpha )} 
          \left( {\frac{1}{{{{\left[ {L_o - (1 - \alpha )s_o} \right]}^2}}} - \frac{1}{{{L_o^2}}}} 
          \right) \frac{M}{EI}  , \qquad \alpha \ne 1   \\
    \end{array}
    \right.
\end{equation}

When rolling of a straight beam with intial shape
\begin{equation}
    \left\{
\begin{array}{l}
    x_o(s_o) = s_o \\
    y_o(s_o) = 0 \\
\end{array}, \qquad 0 \le s_o \le L_o
\right.
\end{equation}
The axial and transverse displacements are,
\begin{equation}
\begin{array}{rl}
    & u(s_o) = \int_0^{s_o} \cos( \theta (s)) ds - s_o \\
    & v(s_o) = \int_0^{s_o} \sin( \theta (s)) ds \\
\end{array}
\end{equation}

Unrolling of an open circular ring with intial shape,
\begin{equation}
    \left\{
\begin{array}{l}
    x_o(s_o) = R \sin(s_o /R)  \\
    y_o(s_o) =  R( 1 - \cos(s_o/R) ) \\
\end{array}, \qquad 0 \le s_o \le L_o
\right.
\end{equation}
The axial and transverse displacements are,
\begin{equation}
\begin{array}{rl}
    & u(s_o) =\int_0^{s_o} \cos( {s}/{R} + \theta (s)) ds - R \sin(s_o /R) \\
    & v(s_o) = \int_0^{s_o} \sin( {s}/{R} + \theta (s)) ds - R( 1 - \cos(s_o/R) )\\
\end{array}
\end{equation}

\newpage

\begin{table}[htbp]
  \centering
  \caption{Load values at the two limit points on the load-displacement curve for pinned-fixed deep arch subjected to a concentrated force.}
    \begin{tabular}{lllccc}
    \toprule
    & Single-patch    & Two-patch     &{ Saje et al.\cite{saje_kinematically_1998}} & Zupan et al.\cite{zupan_finite-element_2003} & Da Deppo et al. \cite{dadeppo_instability_1975}\\
    \midrule
    B  & 897.279 & 897.280 & 897.29 & 897.50 & 897.00 \\
    C  & -73.045   &-73.045   &  &   &  \\
    
    \bottomrule
    \end{tabular}%
  \label{tab:deep}%
\end{table}%

\bigskip

\bigskip

\bigskip

\begin{table}[htbp]
  \centering
  \caption{The first five natural frequencies of the spiral beam (Hz)}
    \begin{tabular}{cccc}
    \toprule
Mode & Present & ABAQUS \\
    \midrule
    1     & 8.544  & 8.545 \\
    2     & 17.097 & 17.099 \\
    3     & 18.798 & 18.800 \\
    4     & 30.867  & 30.871 \\
    5     & 45.958  & 45.966 \\
    \bottomrule
    \end{tabular}%
  \label{tab:spiral_frequencies}%
\end{table}%

\begin{figure}
	\centering
	\includegraphics[width=5in]{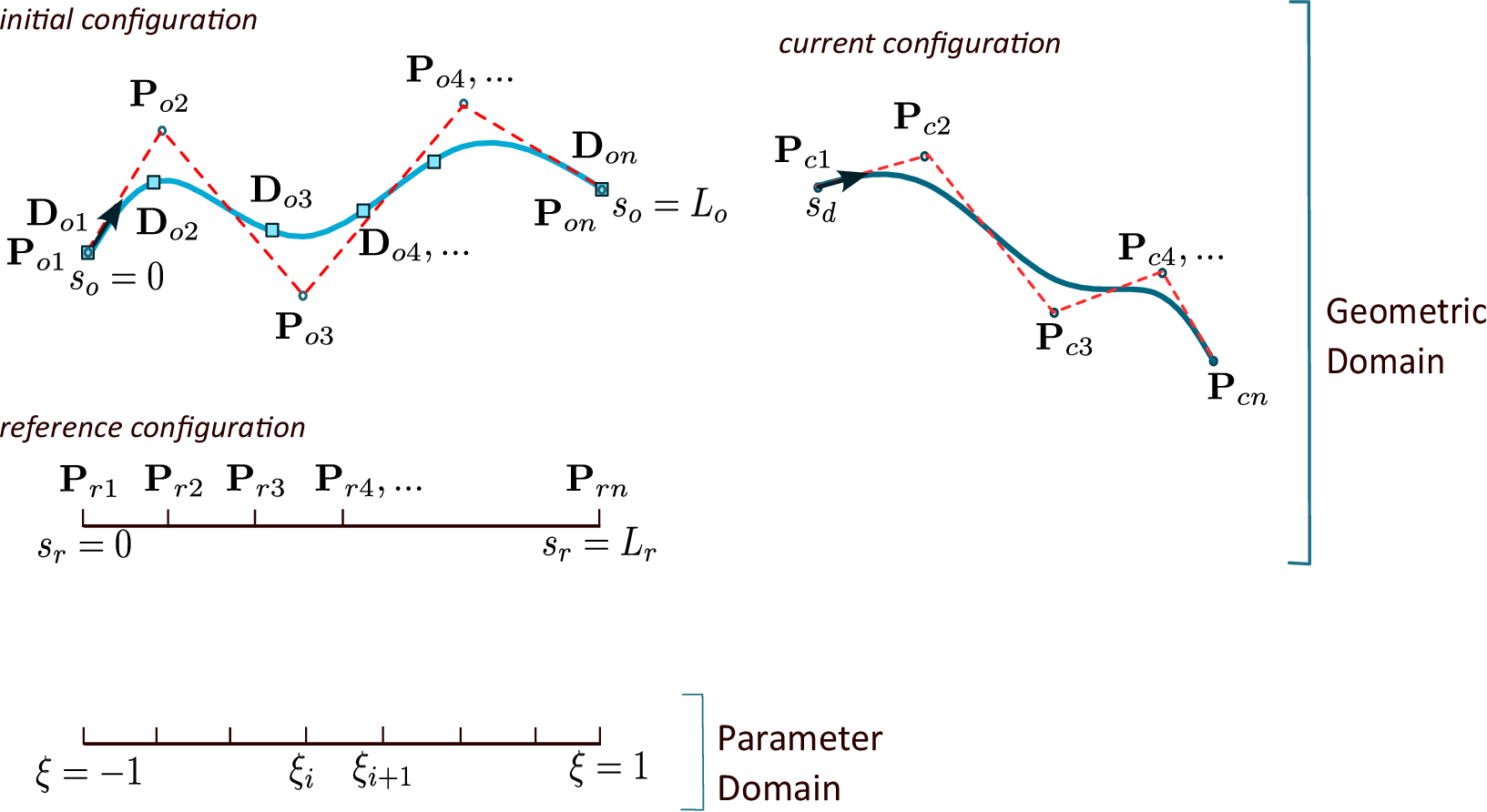}
	\caption{Schematic of the NURBS-based beam representations of the reference, initial, and current beam configurations.}
	\label{fig:transform}
\end{figure}

\begin{figure}[htp]
		\centering
		\subfloat[]{
        \hspace{-0.7in}
			\includegraphics[width=2.5in]{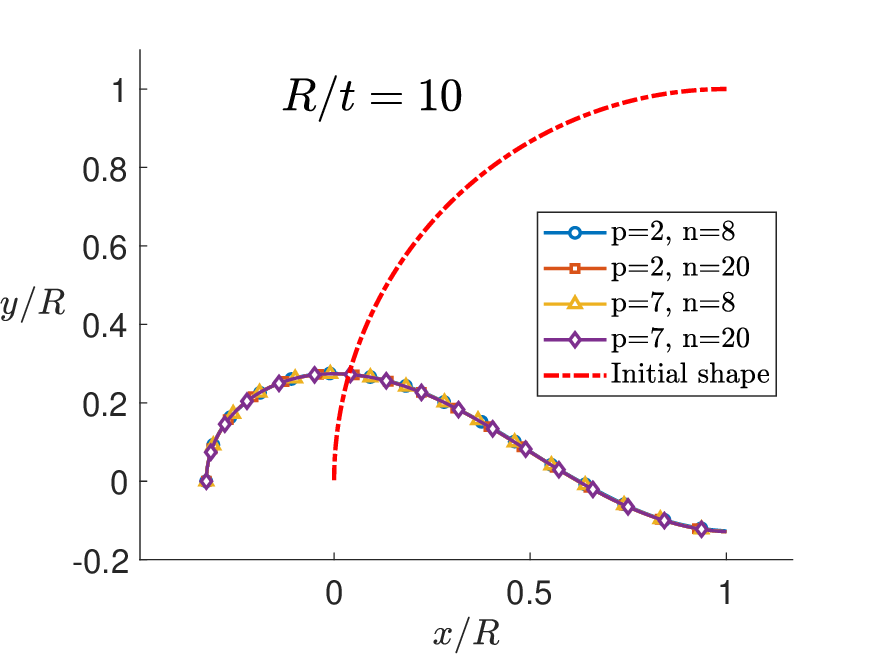}
            \label{fig:ring_compression_Rt10}
		}
		\subfloat[]{
			\hspace{-0.2in}
			\includegraphics[width=2.5in]{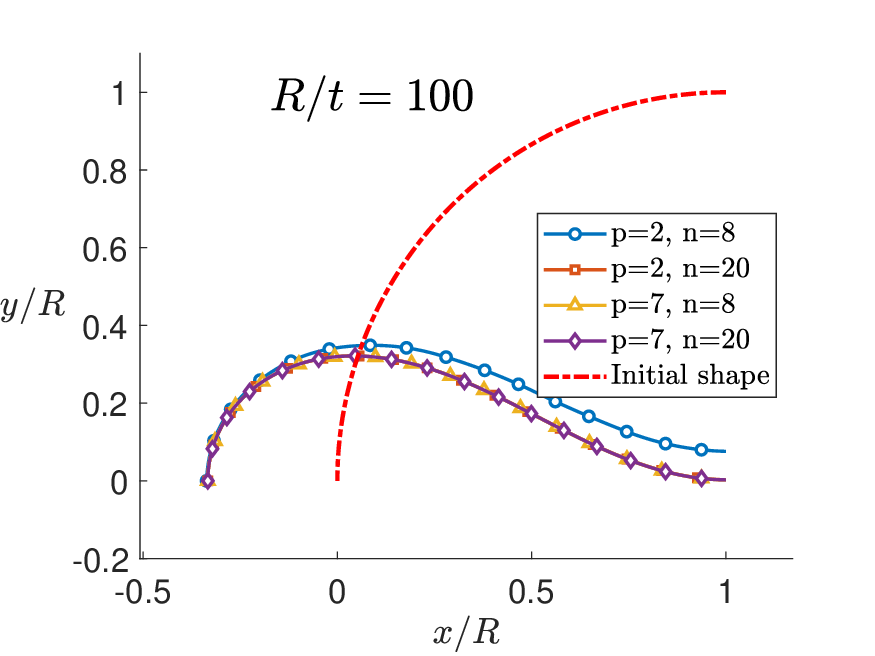}
            \label{fig:ring_compression_Rt100}
		}
        \subfloat[]{
        \hspace{-0.2in}
			\includegraphics[width=2.5in]{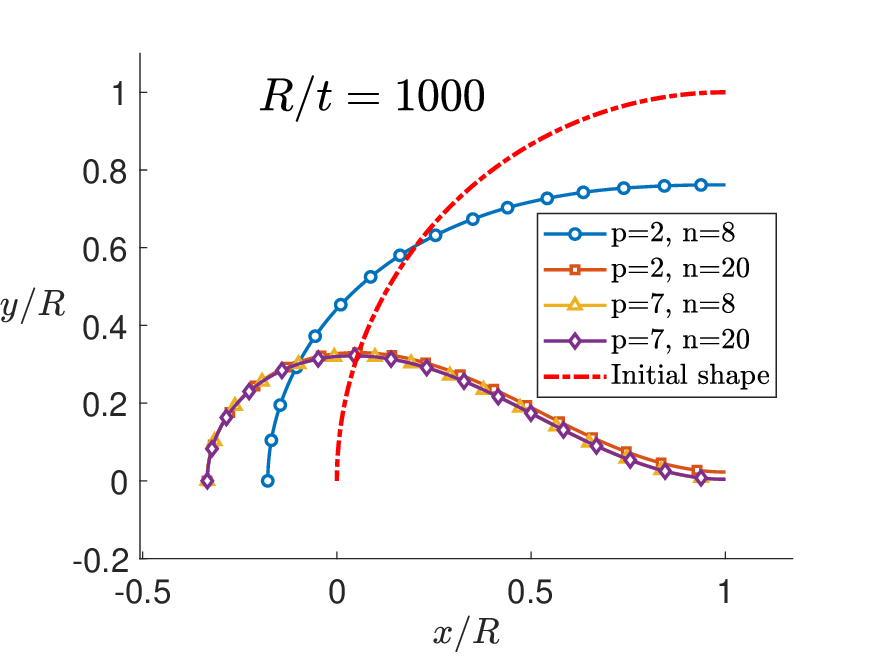}
            \label{fig:ring_compression_Rt1000}
		}\\
        \subfloat[]{
        \hspace{-0.7in}
			\includegraphics[width=2.5in]{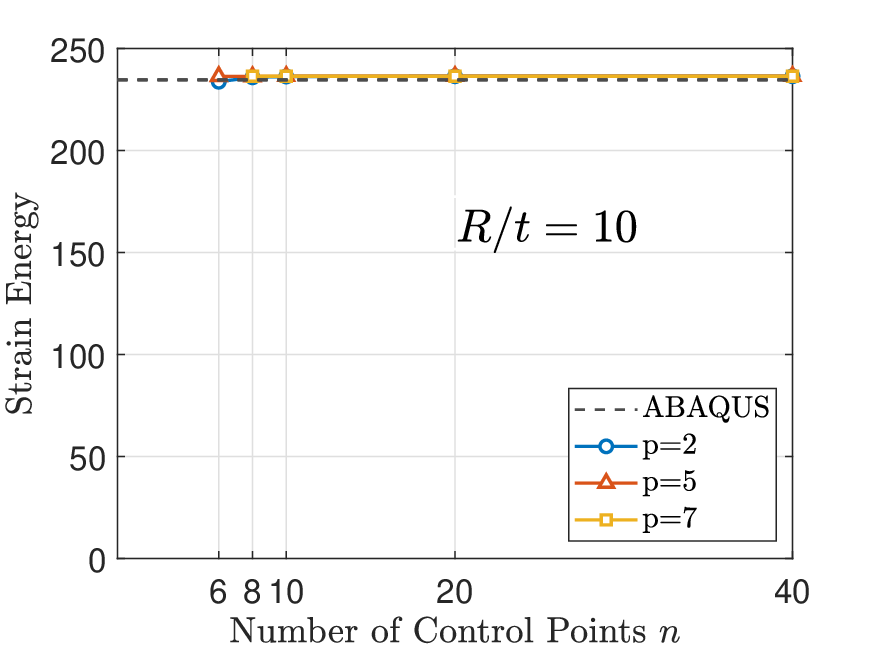}
            \label{fig:ring_compression_Rt10_energy}
		}
		\subfloat[]{
			\hspace{-0.2in}
			\includegraphics[width=2.5in]{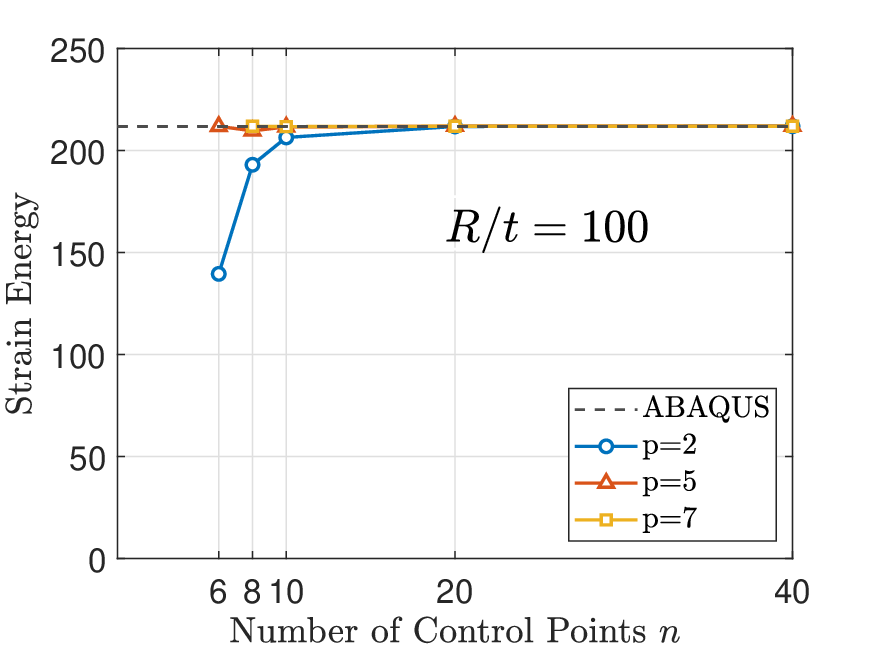}
            \label{fig:ring_compression_Rt100_energy}
		}
        \subfloat[]{
        \hspace{-0.2in}
			\includegraphics[width=2.5in]{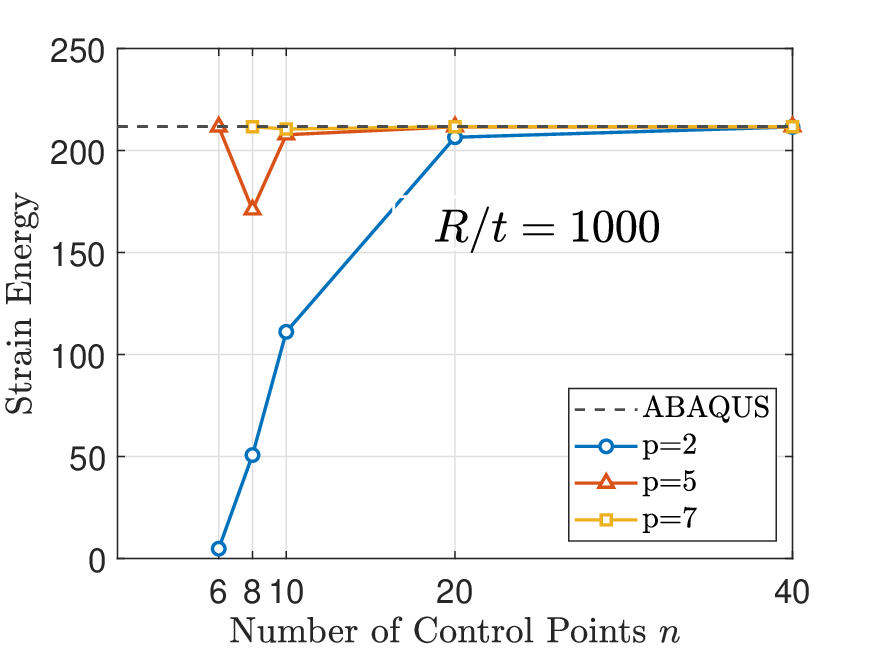}
            \label{fig:ring_compression_Rt1000_energy}
		}

		\caption{ Circular ring subjected to opposite compressive forces $P = 800$. (a)-(c): deformed shape deformed shapes using selected mesh; (d)-(f): total strain energy, for $R/t = 10$, $R/t = 100$, and $R/t = 1000$, respectively.
        }
		\label{fig:ring_compression_multiple_Rt}
	\end{figure}

\begin{figure}[htp]
		\centering
		\subfloat[]{
        \hspace{-0.7in}
			\includegraphics[width=2.5in]{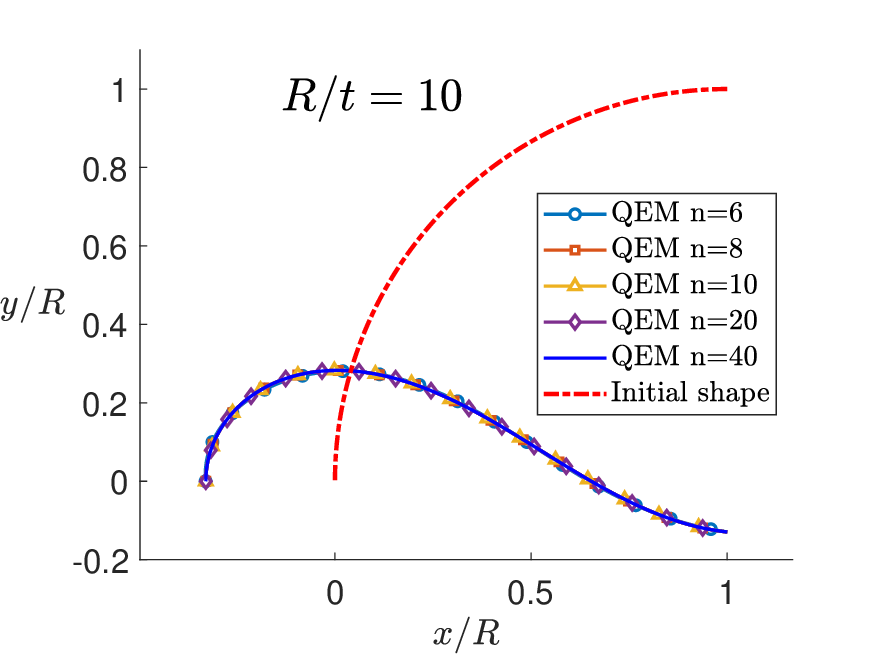}
            \label{fig:ring_compression_Rt10}
		}
		\subfloat[]{
			\hspace{-0.2in}
			\includegraphics[width=2.5in]{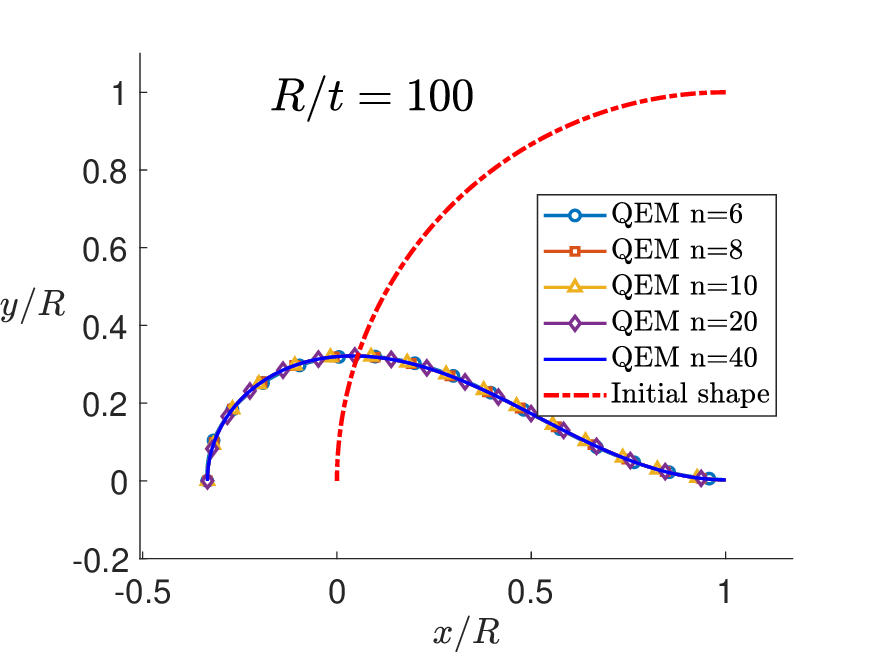}
            \label{fig:ring_compression_Rt100}
		}
        \subfloat[]{
        \hspace{-0.2in}
			\includegraphics[width=2.5in]{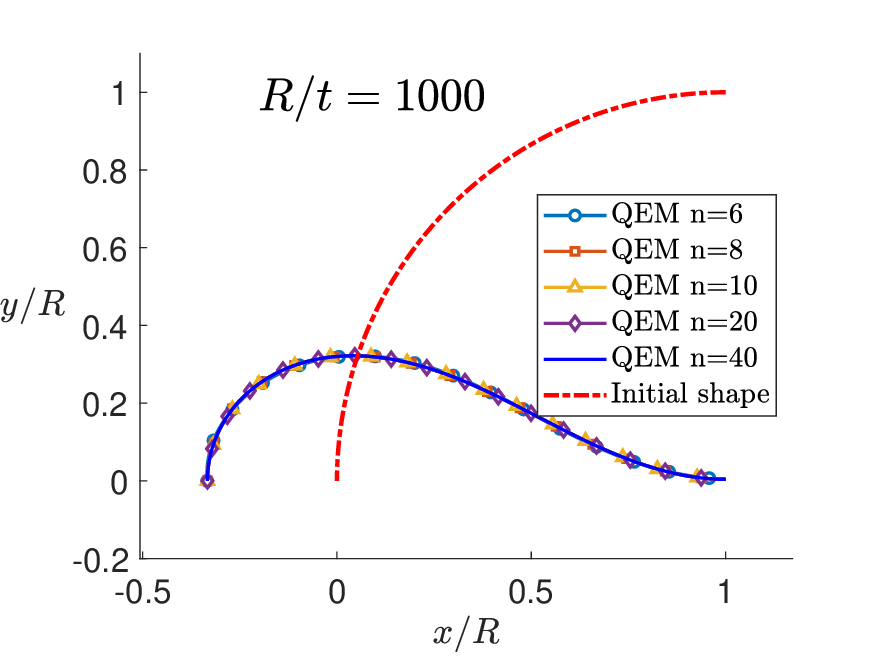}
            \label{fig:ring_compression_Rt1000}
		}\\
        \subfloat[]{
        \hspace{-0.7in}
			\includegraphics[width=2.5in]{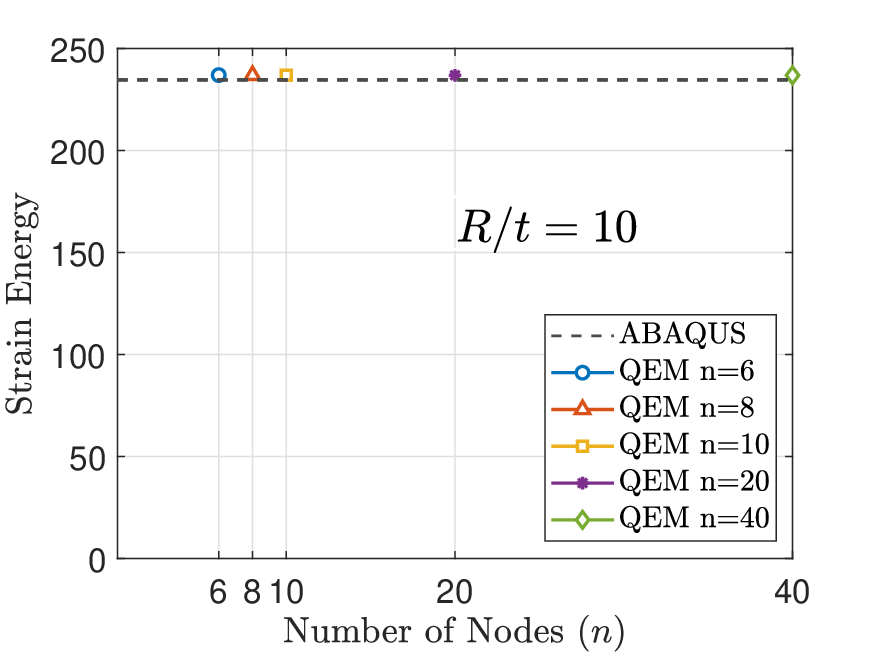}
            \label{fig:ring_compression_Rt10_energy}
		}
		\subfloat[]{
			\hspace{-0.2in}
			\includegraphics[width=2.5in]{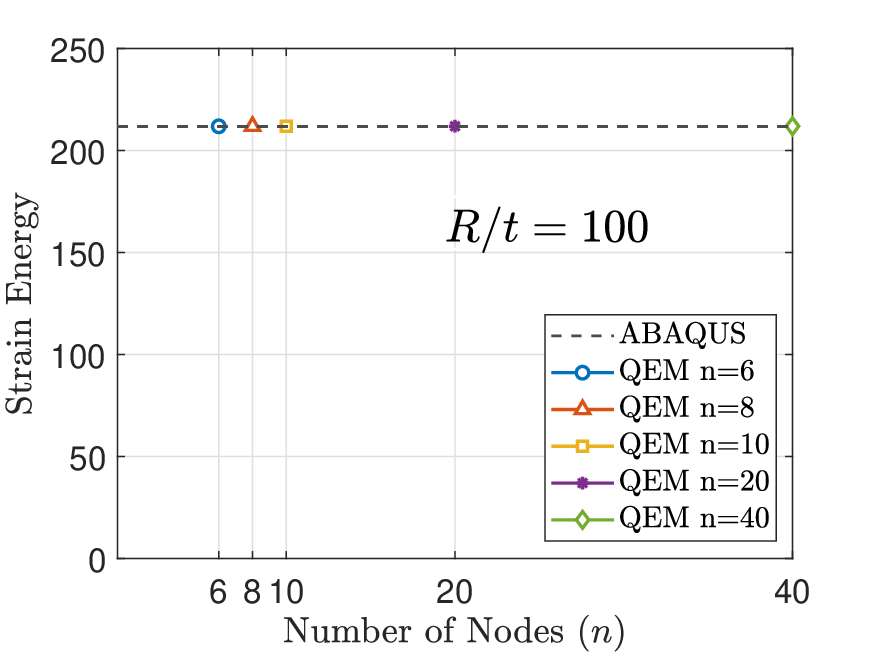}
            \label{fig:ring_compression_Rt100_energy}
		}
        \subfloat[]{
        \hspace{-0.2in}
			\includegraphics[width=2.5in]{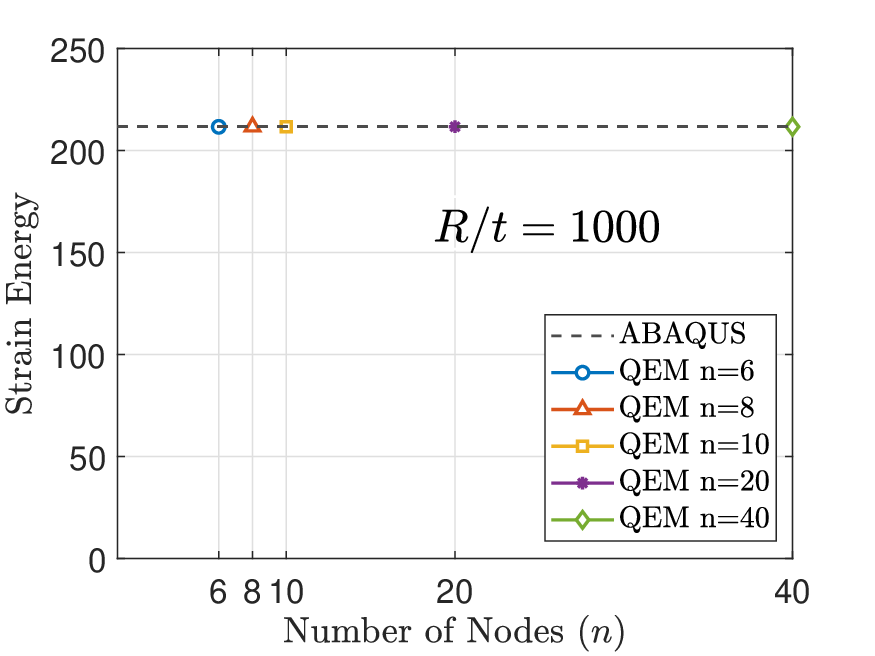}
            \label{fig:ring_compression_Rt1000_energy}
		}

		\caption{ Circular ring subjected to opposite compressive forces $P = 800$. (a)-(c): deformed shape deformed shapes using selected mesh; (d)-(f): total strain energy, for $R/t = 10$, $R/t = 100$, and $R/t = 1000$, respectively.
        }
		\label{fig:ring_compression_multiple_Rt}
	\end{figure}

	
	\begin{figure}
		\centering
		\includegraphics[width=3.in]{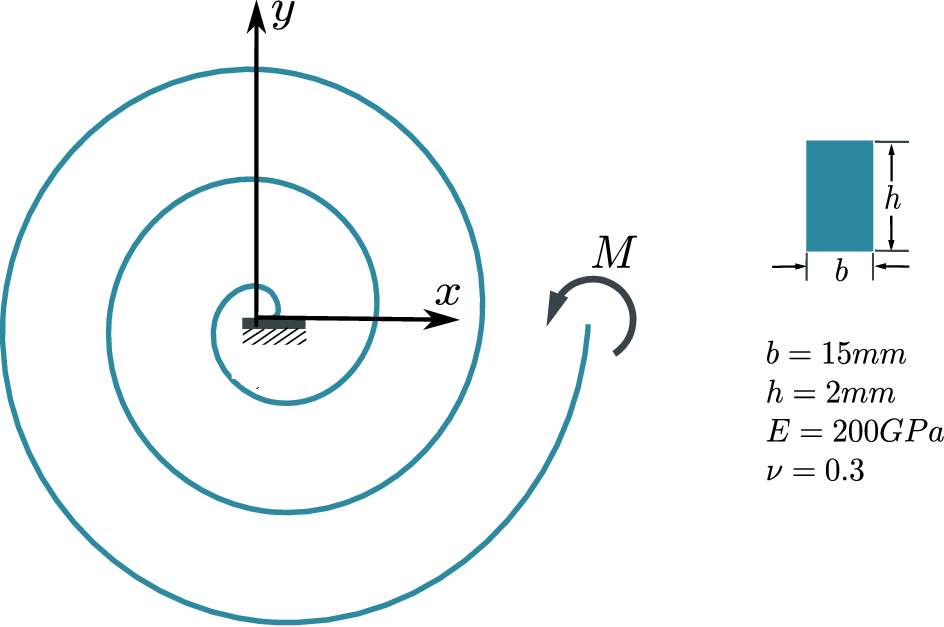}
		\caption{Schematic of a spiral beam subjected to an applied bending moment. }
		\label{fig:Spiral}
	\end{figure}

\end{document}